\documentclass[journal]{IEEEtran}

\usepackage{amsmath,amssymb,amsfonts}
\usepackage{mathrsfs}
\usepackage{dsfont}
\usepackage{pifont}         
\usepackage{nicefrac}       

\usepackage[table,xcdraw]{xcolor} 
\usepackage{graphicx}
\usepackage{tikz}

\usepackage{array}
\usepackage{booktabs}       
\usepackage{multirow}       
\usepackage{multicol}       
\usepackage{tabularx}       
\usepackage{makecell}       
\usepackage{arydshln}       

\usepackage{algorithm}
\usepackage{algorithmic}
\usepackage[font=small]{caption}

\usepackage{cite}           
\usepackage{url}
\usepackage{stfloats}       
\usepackage{microtype}      
\usepackage[normalem]{ulem} 

\usepackage{subfig}         
\usepackage{capt-of}        

\usepackage[colorlinks=true, citecolor=green, linkcolor=red, urlcolor=blue]{hyperref}

\definecolor{mygray}{gray}{0.9}
\definecolor{mygray1}{rgb}{0.9, 0.9, 0.9}

\newcommand{\cmark}{\ding{51}}  
\newcommand{\xmark}{\ding{55}}  

\newcommand\hl{\bgroup\markoverwith
	{\textcolor{yellow}{\rule[-.5ex]{2pt}{2.5ex}}}\ULon}

\newcommand{\safeincludegraphics}[2][]{%
  \IfFileExists{#2}{\includegraphics[#1]{#2}}{%
    \fbox{\parbox[c][2.2in][c]{0.95\linewidth}{\centering\scriptsize Missing figure file\\\texttt{\detokenize{#2}}}}%
  }%
}

\definecolor{mygray}{gray}{0.9}
\definecolor{mygray1}{rgb}{0.9, 0.9, 0.9}

\usepackage[most]{tcolorbox}
\usepackage{listings}
\usepackage{xcolor}

\lstdefinestyle{promptstyle}{
  basicstyle=\ttfamily\fontsize{7pt}{5pt}\selectfont,
  columns=fullflexible,
  keepspaces=true,
  breaklines=true,
  breakatwhitespace=false,
  showstringspaces=false,
  escapeinside={(*@}{@*)},
}

\newtcblisting{promptbox}[2][]{
  enhanced,
  breakable,
  colback=black!5,
  colframe=black!70,
  boxrule=0.4pt,
  arc=1mm,
  left=1mm,
  right=1mm,
  top=1mm,
  bottom=1mm,
  listing only,
  listing options={style=promptstyle},
  title={#2},
  fonttitle=\small\bfseries,
  #1
}
\allowdisplaybreaks

\begin{document}

\title{IMAGAgent: Orchestrating Multi-Turn Image Editing via Constraint-Aware Planning and Reflection}

\author{Fei Shen, Chengyu Xie, Lihong Wang, Zhanyi Zhang, Xin Jiang, \\ 
Xiaoyu Du, Jinhui Tang, \IEEEmembership{Senior Member, IEEE}, and Tat-Seng Chua

\thanks{This work is supported by the Jiangsu Provincial Science and Technology Major Project (No. BG2024042).(Corresponding author: Jinhui Tang.)}
\thanks{Fei Shen, and Tat-Seng Chua are with the NExT++ Research Centre, National University of Singapore, Singapore, (e-mail: \{shenfei29, dcscts\}@nus.edu.sg).}
\thanks{Chengyu Xie, Lihong Wang, Zhanyi Zhang, Xin Jiang and Xiaoyu Du are with the School of Computer Science and Engineering, Nanjing University of Science and Technology, Nanjing, China. E-mail: \{xiechengyu, lihongwang, zhanyizhang, xinjiang, duxy\}@njust.edu.cn.}
\thanks{Jinhui Tang is with the College of Information Science and Technology and
Artificial Intelligence, Nanjing Forestry University, Nanjing 210037, China. E-mail: jinhuitang@njfu.edu.cn.}
}

\markboth{IEEE Transactions on Multimedia, Vol. xx, No. xx}
{Shen \MakeLowercase{\textit{et al.}}: IMAGAgent: Orchestrating Multi-Turn Image Editing via Constraint-Aware Planning and Reflection}

\maketitle

\begin{abstract}
Existing multi-turn image editing paradigms are often confined to isolated single-step execution. 
Due to a lack of context-awareness and closed-loop feedback mechanisms, they are prone to error accumulation and semantic drift during multi-turn interactions, ultimately resulting in severe structural distortion of the generated images.
To address this, we propose IMAGAgent, a multi-turn image editing agent framework based on a ``plan--execute--reflect'' closed-loop mechanism that achieves deep synergy among instruction parsing, tool scheduling, and adaptive correction within a unified pipeline.
Specifically, we first present a constraint-aware planning module that leverages a vision-language model (VLM) to precisely decompose complex natural language instructions into a series of executable sub-tasks, governed by target singularity, semantic atomicity, and visual perceptibility.
Then, the tool-chain orchestration module dynamically constructs execution paths based on the current image, the current sub-task, and the historical context, enabling adaptive scheduling and collaborative operation among heterogeneous operation models covering image retrieval, segmentation, detection, and editing.
Finally, we devise a multi-expert collaborative reflection mechanism in which a central large language model (LLM) synthesizes evaluations from multiple vision-language models into holistic feedback, triggering fine-grained self-correction and recording feedback outcomes to optimize future decisions.
Extensive experiments on our constructed MTEditBench and the MagicBrush dataset demonstrate that IMAGAgent achieves performance significantly superior to existing methods in terms of instruction consistency, editing precision, and overall quality.
The code is available at \url{https://github.com/hackermmzz/IMAGAgent}.
\end{abstract}


\section{Introduction}
\IEEEPARstart{M}{ulti-turn} image editing~\cite{cheng2020sequential,ye2025imgedit,zhang2025context} is becoming increasingly important in practical creative workflows, as it enables users to progressively refine visual content through a sequence of natural language instructions rather than relying on a single, all-in-one editing request. Although diffusion models~\cite{suvorov2022resolution,lugmayr2022repaint} and tool-augmented agents~\cite{shen2023hugginggpt,yang2023mm,wu2023visual} have established strong foundations for image editing by providing powerful generation and task coordination capabilities, maintaining semantic and structural stability across multiple editing turns remains a critical and largely unresolved bottleneck. The core challenge lies in satisfying the current editing intent while rigorously preserving the semantic and structural consistency established by the editing history.

\begin{figure}[t]
    \centering
    \includegraphics[width=0.9\linewidth]{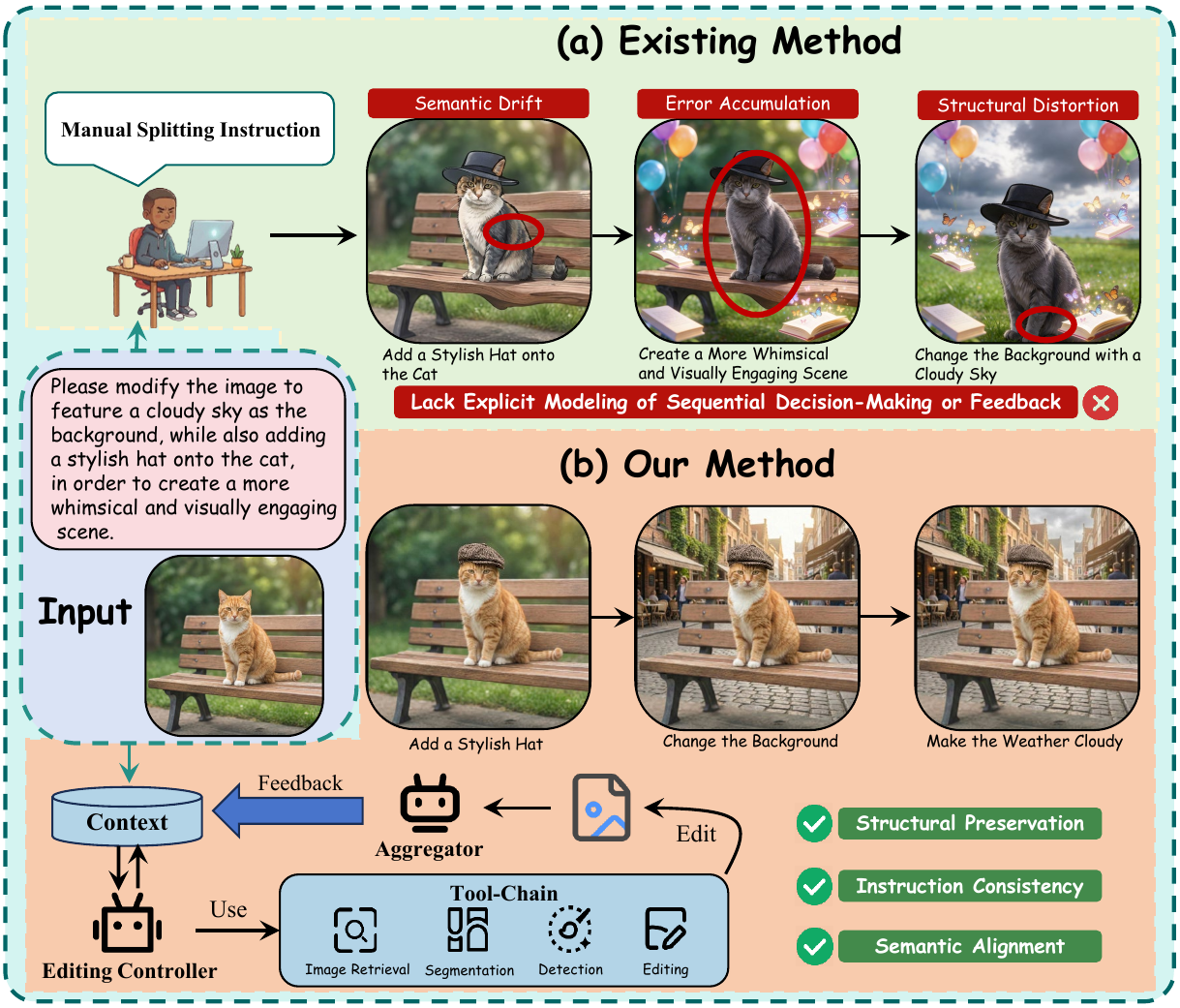}
    \caption{(a) \textbf{Existing methods} suffer from error accumulation due to isolated execution. (b) \textbf{IMAGAgent} ensures structural and intentional consistency through closed-loop feedback.}
    \vspace{-0.5cm}
    \label{fig:motivation}
\end{figure}

Unlike single-turn editing~\cite{meng2021sdedit,couairon2022diffedit}, the core challenge in multi-turn scenarios is maintaining stability. Since each intermediate output becomes the next input, errors easily accumulate and destabilize the editing trajectory (Fig.~\ref{fig:motivation}). Existing approaches often lack explicit mechanisms to enforce cross-turn constraints~\cite{wang2025complexbench}, leading to three recurring failure modes. First, error accumulation occurs as uncorrected deviations propagate. As shown in Fig.~\ref{fig:motivation} (a), early flaws amplify during the ``whimsical scene'' edit, introducing artifacts and blurring the subject. Second, semantic drift emerges when new edits overwrite established content; for example, adding the hat unintentionally alters the subject's original identity. Third, structural distortion arises when local modifications disrupt global coherence. The final background change inadvertently distorts both the subject and the underlying bench's geometry.

Existing approaches generally lack a closed-loop decision mechanism explicitly designed for cross-turn constraint maintenance. Single-turn diffusion-based editing~\cite{meng2021sdedit,couairon2022diffedit} and controllable generation methods~\cite{zhang2023adding,mou2024t2i,brooks2023instructpix2pix} typically operate under a static "one instruction, one edit" regime, lacking explicit mechanisms for sequential decision-making. While recent multi-turn studies~\cite{joseph2024iterative,yin2025reasonedit,shi2021learning} attempt to mitigate degradation via iterative strategies, they often rely on rigid workflows and cannot adaptively adjust actions based on intermediate outcomes. Although tool-augmented vision agents~\cite{shen2023hugginggpt,yang2023mm} can coordinate heterogeneous multimodal models for complex visual tasks, they typically do not evaluate intermediate editing results or adapt subsequent actions accordingly. As a result, they remain vulnerable to semantic drift and structural degradation in long-horizon image editing.

To bridge these gaps, we propose IMAGAgent, a multi-turn image editing agent built upon a closed-loop ``plan--execute--reflect'' mechanism. Its key distinction lies in history-aware constraint maintenance for long-horizon editing, rather than simple tool combination or undirected retry. IMAGAgent comprises three modules:
(1) a constraint-aware planning module, which utilizes vision-language models (VLMs) to precisely decompose complex instructions into atomic sub-tasks, strictly adhering to constraints of target singularity, semantic atomicity, and visual perceptibility; 
(2) a tool-chain orchestration module, which dynamically constructs execution paths based on the current image, current sub-task, and historical context, enabling adaptive scheduling across heterogeneous expert models for retrieval, segmentation, detection, and editing; 
and (3) a multi-expert collaborative reflection mechanism, which employs a panel of VLMs to provide multi-dimensional feedback. This holistic critique not only automatically triggers self-correction for erroneous iterations but also records feedback outcomes into the historical context to optimize subsequent decisions.
In addition, we construct MTEditBench, a comprehensive benchmark specifically designed for long-horizon multi-turn image editing. 
Building upon this new benchmark and the MagicBrush dataset, we conduct systematic evaluations to more comprehensively analyze the system's overall performance across varying editing turns and complexities.
Our main contributions are summarized as follows:

\begin{itemize}
  
    \item We formulate long-horizon multi-turn image editing as a history-aware cross-turn constraint-maintenance problem and propose IMAGAgent, which explicitly records completed edits and failure feedback to preserve prior requirements across subsequent turns.
    \item We present a constraint-aware planning module and a tool-chain orchestration module, which together ensure that each sub-task is executable under explicit visual and semantic constraints, enabling adaptive scheduling across heterogeneous vision tools.
    \item We design a multi-expert collaborative reflection mechanism that transforms visual feedback into actionable correction signals, effectively suppressing error accumulation and semantic drift in long-horizon editing.
    \item We conduct extensive experiments on our constructed MTEditBench and the MagicBrush dataset, providing a systematic assessment of long-horizon multi-turn image editing and showing that IMAGAgent significantly outperforms current SOTA baselines.
\end{itemize}

\begin{figure*}[t]
    \centering
    \includegraphics[width=0.98\linewidth]{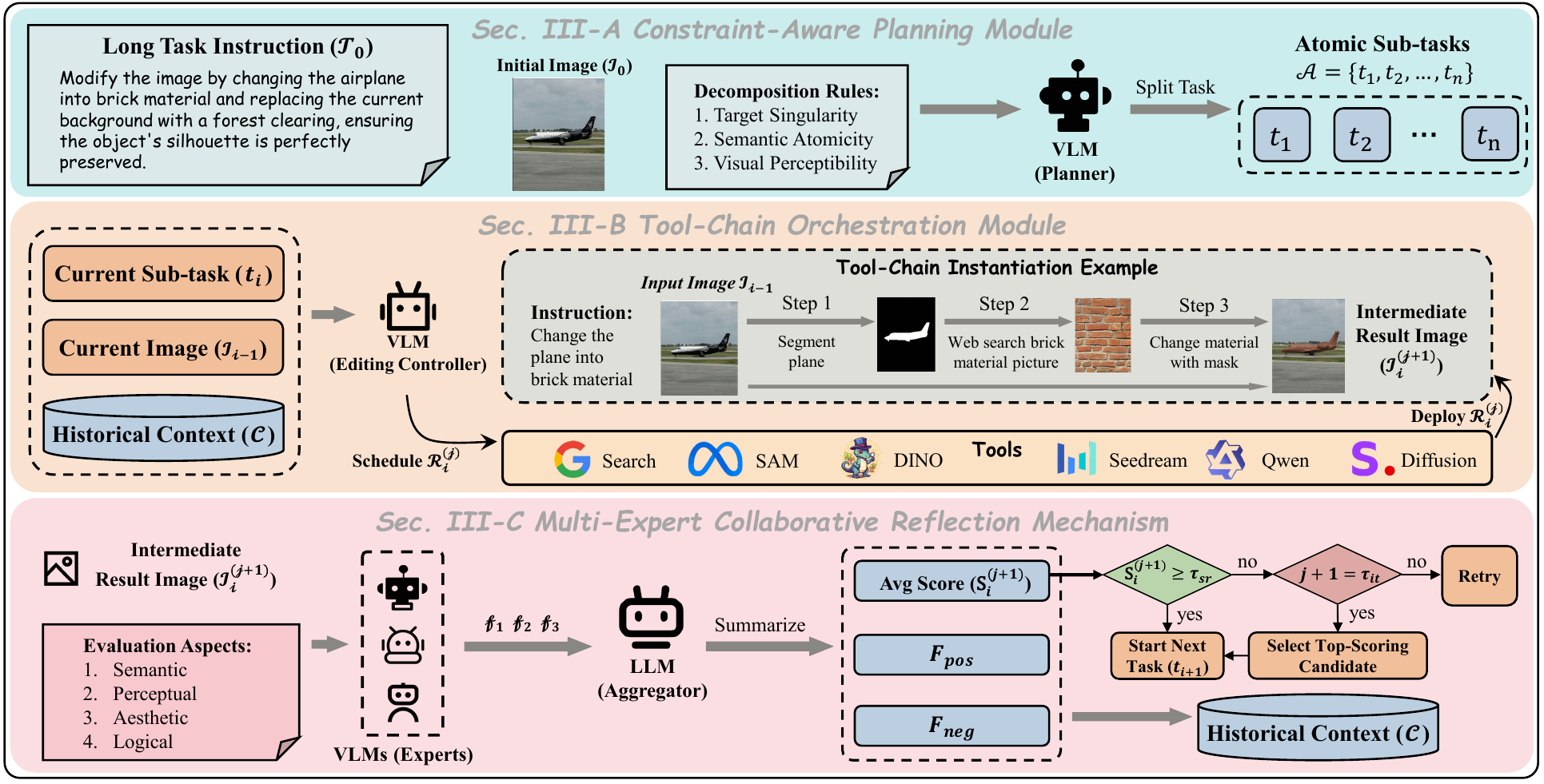}
    \caption{\textbf{IMAGAgent Framework.} We reformulate multi-turn image editing into a closed-loop process with three modules: (1) constraint-aware planning for VLM-based instruction decomposition; (2) tool-chain orchestration for dynamic scheduling of vision tools; and (3) multi-expert collaborative reflection for fine-grained critique and self-correction.}
    \label{fig:framework}
    \vspace{-0.3cm}
\end{figure*}

\section{Related Work}~\label{sec:rw}

\noindent\textbf{From Single-Turn to Multi-Turn Image Editing.}
Early image editing methods~\cite{zhao2024ultraedit,mao2025visual} primarily focused on single-turn generation quality and local controllability. For example, SDEdit~\cite{meng2021sdedit} formulates editing as user-guided stochastic denoising, while Prompt-to-Prompt~\cite{hertz2022prompt} controls cross-attention to preserve spatial layouts. More recent approaches improve instruction following through large-scale training and specialized architectures: UltraEdit~\cite{zhao2024ultraedit} employs automated region-aware supervision, and VAREdit~\cite{mao2025visual} adopts visual autoregressive modeling. Unified frameworks such as OmniGen~\cite{xiao2025omnigen} and ACE++~\cite{mao2025ace++} further support diverse conditional generation tasks, while other studies preserve editing history and background consistency during iterative interactions~\cite{kim2025improving}. 
However, these methods~\cite{zhao2024ultraedit,mao2025visual,xiao2025omnigen} remain largely optimized for isolated or weakly coupled edits, without explicit mechanisms for modeling cross-turn dependencies, enforcing accumulated constraints, or recovering from intermediate failures.
In multi-turn settings, the primary challenge shifts toward maintaining semantic and structural consistency across dependent edits. VINCIE~\cite{qu2025vincie} learns contextual editing from video-derived multimodal sequences, while ICEdit~\cite{zhang2025context} strengthens in-context instruction execution with diffusion transformers. Although these methods improve sequential continuity, they do not explicitly diagnose or correct intermediate failures and therefore remain susceptible to error accumulation and semantic drift over long editing horizons.

\noindent\textbf{Multimodal Agents with Vision-Language Feedback.}
Multimodal agents~\cite{gupta2023visual,liu2024llava,wang2024genartist} integrate perception, reasoning, and tool orchestration for increasingly complex real-world visual tasks. 
Visual ChatGPT~\cite{wu2023visual} coordinates multiple visual foundation models for interactive editing, while ViperGPT~\cite{suris2023vipergpt} and GPT4Tools~\cite{yang2023gpt4tools} generate executable programs for invoking multimodal tools. More recent systems incorporate intermediate verification: GenArtist~\cite{wang2024genartist} performs request decomposition and step-wise checking, and FaST~\cite{sun2024visual} adapts its reasoning strategy to task uncertainty. Nevertheless, these general-purpose agents typically lack explicit mechanisms to represent and encourage preservation of prior edits accumulated from earlier edits, making it difficult to preserve previously satisfied content, spatial, and structural requirements over long editing horizons.
Another line of work~\cite{yue2024mllm,wu2025gui,shinn2023reflexion} explores memory, reflection, and recovery in long-horizon agent trajectories. MART~\cite{yue2024mllm} retrieves historical trajectories to guide future decisions, while OpenCUA~\cite{wang2025opencua} improves robustness through reflective reasoning and failure recovery. GUI-Reflection~\cite{wu2025gui} and Reflexion~\cite{shinn2023reflexion} further demonstrate the benefits of explicit self-correction and verbal memory. However, these mechanisms are generally designed for task-level reasoning and recovery, rather than for jointly evaluating instruction compliance, visual fidelity, and the preservation of accumulated constraints after each editing step. IMAGAgent adapts reflection to multi-turn image editing by more tightly coupling constraint-aware planning and history-conditioned tool orchestration with multi-expert visual feedback that guides subsequent corrections.

\section{Method}~\label{sec:method}

\noindent\textbf{Problem Formulation.}
We formalize multi-turn image editing as a sequential optimization process driven by feedback. 
Let $\mathcal{I}_0$ denote the initial image and $\mathcal{T}_0$ represent the complex user instruction. 
Our objective is to generate a stable edited image $\mathcal{I}^*$ via a mapping function $\Phi$, governed by a closed-loop pipeline comprising the constraint-aware planning module, tool-chain orchestration module, and multi-expert collaborative reflection mechanism:
\begin{equation}
    \mathcal{I}^* = \Phi(\mathcal{I}_0, \mathcal{T}_0),
    \label{eq:problem_formulation}
\end{equation}
where the process keeps a dynamic historical context $\mathcal{C}$ to facilitate sequential reasoning across multiple turns.
To ensure continuity, $\mathcal{C}$ is defined as the union of the initial visual state and the interaction trajectory accumulated from subsequent turns: $\mathcal{C} = \{ \mathcal{R}_k, \mathcal{I}_k, \mathcal{F}_k \}_{k \ge 1} \cup \{ \mathcal{I}_0 \}$. 
Here, $k$ indexes the completed editing attempts, associating each attempt with its executed orchestration plan $\mathcal{R}_k$ (reasoning process, tool-chain, and parameters), resulting visual state $\mathcal{I}_k$, and critique feedback $\mathcal{F}_k$. 
Together, these elements allow the controller to trace both successful modifications and unsuccessful attempts throughout the editing trajectory. 
Rather than serving as a passive execution log, $\mathcal{C}$ functions as a history-conditioned trajectory memory. Accepted visual states identify the completed modifications that should be preserved, while critique feedback records observed failures and correction cues that should influence subsequent decisions. In this way, the current editing objective is optimized jointly with the constraints accumulated from previous turns.

\noindent\textbf{Framework Overview.}
As shown in Fig.~\ref{fig:framework}, IMAGAgent operates via a closed-loop pipeline comprising three core modules: the constraint-aware planning module, tool-chain orchestration module, and multi-expert collaborative reflection mechanism. 
The workflow initiates with the constraint-aware planning module, which leverages a VLM to ground high-level instructions into the visual context and decompose them into a sequence of atomic sub-tasks. These sub-tasks are processed by the tool-chain orchestration module, which dynamically schedules heterogeneous tool modules to execute precise edits.
To ensure semantic fidelity and structural stability, the multi-expert collaborative reflection mechanism evaluates intermediate outputs, providing fine-grained feedback that triggers self-correction loops before proceeding to the next sub-task. For each sub-task, a candidate satisfying the acceptance threshold is accepted as the input to the next sub-task; otherwise, if the attempt budget is exhausted without a candidate meeting the acceptance threshold, the highest-scoring candidate is selected as a best-effort output for the current sub-task.
This critique-guided refinement aims to limit error propagation across successive editing turns.
The complete inference procedure is detailed in Algorithm~\ref{alg:imagagent}.

\begin{algorithm}[t]
\caption{IMAGAgent Inference}
\label{alg:imagagent}
\begin{flushleft}
\textbf{Input:} Initial image $\mathcal{I}_0$, instruction $\mathcal{T}_0$ \\
\textbf{Parameters:} Acceptance threshold $\tau_{sr}$,
maximum editing attempts $\tau_{it}$ \\
\textbf{Output:} Final edited image $\mathcal{I}^*$
\end{flushleft}
\begin{algorithmic}[1]
\STATE $\mathcal{C} \leftarrow \{\mathcal{I}_0\}$
\STATE \textcolor{gray}{\textit{// Constraint-Aware Planning}}
\STATE $\mathcal{A} \leftarrow \Psi(\mathcal{I}_0, \mathcal{T}_0)$
\STATE Consolidate and reorder $\mathcal{A}$ by semantic dependencies
\STATE $n \leftarrow |\mathcal{A}|$
\FOR{$i = 1$ \TO $n$}
    \FOR{$j = 0$ \TO $\tau_{it}-1$}
        \STATE \textcolor{gray}{\textit{// Tool-Chain Orchestration}}
        \STATE $\mathcal{R}_i^{(j)} \leftarrow
        \text{Agent}(\mathcal{I}_{i-1}, t_i, \mathcal{C})$
        \STATE $\mathcal{I}_i^{(j+1)} \leftarrow
        \text{Execute}(\mathcal{R}_i^{(j)}, \mathcal{I}_{i-1})$
        \STATE \textcolor{gray}{\textit{// Multi-Expert Reflection}}
        \STATE Obtain expert critiques $\{f_k\}_{k=1}^{3}$
        for $\mathcal{I}_i^{(j+1)}$
        \STATE $\mathcal{F}_i^{(j+1)} \leftarrow
        \text{Aggregator}(\{f_k\}_{k=1}^{3})$
        \STATE Extract $S_i^{(j+1)}$ from $\mathcal{F}_i^{(j+1)}$
        \STATE \textcolor{gray}{\textit{// Retain the current attempt}}
        \STATE $\mathcal{C} \leftarrow \mathcal{C} \cup
        \{\mathcal{R}_i^{(j)}, \mathcal{I}_i^{(j+1)},
        \mathcal{F}_i^{(j+1)}\}$
        \IF{$S_i^{(j+1)} \ge \tau_{sr}$}
            \STATE \textcolor{gray}{\textit{// Accept the current candidate}}
            \STATE $\mathcal{I}_i \leftarrow \mathcal{I}_i^{(j+1)}$
            \STATE \textbf{break}
        \ELSIF{$j+1 = \tau_{it}$}
            \STATE \textcolor{gray}{\textit{// Best-effort fallback}}
            \STATE $k^\star \leftarrow
            \arg\max_{1 \le k \le \tau_{it}} S_i^{(k)}$
            \STATE $\mathcal{I}_i \leftarrow \mathcal{I}_i^{(k^\star)}$
        \ENDIF
    \ENDFOR
    \STATE \textcolor{gray}{\textit{// Record the selected output}}
    \STATE $\mathcal{C} \leftarrow \mathcal{C} \cup \{\mathcal{I}_i\}$
\ENDFOR
\STATE $\mathcal{I}^* \leftarrow \mathcal{I}_n$
\RETURN $\mathcal{I}^*$
\end{algorithmic}
\end{algorithm}

\subsection{Constraint-Aware Planning}
The constraint-aware planning module maps the potentially ambiguous global instruction $\mathcal{T}_0$ into a structured sequence of atomic sub-tasks $\mathcal{A} = \{t_1, t_2, \dots, t_n\}$. This decomposition is performed by a VLM-based planner $\Psi$:
\begin{equation}
    \mathcal{A} = \Psi(\mathcal{I}_0, \mathcal{T}_0)
    \label{eq:planning_decomposition}.
\end{equation}
Distinct from text-only planners, our VLM planner grounds instruction decomposition in the spatial layout and visual content of $\mathcal{I}_0$. To ensure that each planned step can be reliably executed by downstream tools, every sub-task $t_i$ must satisfy three criteria. First, \textit{target singularity} requires each sub-task to operate on one entity or a visually coherent group, rather than multiple unrelated objects. Second, \textit{semantic atomicity} requires each sub-task to express a single coherent editing objective that should not be meaningfully divided further. Third, \textit{visual perceptibility} requires the action to introduce an observable change in the image, such as modifying an object, attribute, or background.

Instruction decomposition alone is insufficient for edits involving dependencies or redundant operations. The planner therefore merges repeated or chained sub-tasks and reorders the sequence $\mathcal{A}$ according to their semantic dependencies, ensuring that prerequisite edits are completed before dependent ones. This process yields an executable and causally consistent editing sequence.

\subsection{Tool-Chain Orchestration}
Upon establishing the sub-task sequence $\mathcal{A}$, the system enters the execution phase. To facilitate robust decision-making, we utilize a dynamic historical context set $\mathcal{C}$ that encapsulates reasoning trajectories, critique feedback, and intermediate states. For the $i$-th sub-task, the agent operates in a feedback-driven re-execution loop (indexed by $j$), where each iteration corresponds to a new editing attempt conditioned on the accumulated critique feedback.
The agent employs a chain-of-thought (CoT) mechanism~\cite{wei2022chain} to orchestrate heterogeneous tools. Conditioned on the current image state $\mathcal{I}_{i-1}$, the current atomic sub-task $t_i$, and the historical context $\mathcal{C}$, the agent generates a comprehensive orchestration plan $\mathcal{R}_i^{(j)}$, which encapsulates the chain-of-thought rationale, the selected tool-chain, and the optimized execution parameters:
\begin{equation}
    \mathcal{R}_i^{(j)} = \text{Agent}(\mathcal{I}_{i-1}, t_i, \mathcal{C}).
\end{equation}
Consequently, retries are not independent resampling attempts. At each iteration, the controller retrieves previous orchestration decisions, intermediate outputs, and critique feedback from $\mathcal{C}$ to revise the tool-chain and execution parameters, while treating previously accepted edits as constraints to be preserved.
Subsequently, the refined candidate image is generated by executing this plan:
\begin{equation}
    \mathcal{I}_i^{(j+1)} = \text{Execute}(\mathcal{R}_i^{(j)}, \mathcal{I}_{i-1}).
\end{equation}

\subsection{Multi-Expert Collaborative Reflection}
To suppress semantic drift and ensure structural fidelity, we deploy a panel of three distinct VLM experts ($k=1, 2, 3$) to conduct a multi-dimensional assessment. Grounded in a rigorous evaluation rubric covering semantic alignment, perceptual quality, aesthetic assessment, and logical consistency, each expert independently scrutinizes the generated candidate $\mathcal{I}_i^{(j+1)}$ to yield a raw critique tuple $f_k = \{f_{pos}^{(k)}, f_{neg}^{(k)}, s^{(k)}\}$. Here, the components represent the expert's distinct feedback on positive traits, negative defects, and a quantitative score, respectively. Subsequently, to resolve potential conflicts and distill actionable insights, a central LLM aggregator synthesizes the set of individual critiques $\{f_1, f_2, f_3\}$ into a unified and actionable consensus feedback triplet $\mathcal{F}_i^{(j+1)}$. This aggregation process is formulated as:

\begin{equation} \mathcal{F}_i^{(j+1)} = \{F_{pos}, F_{neg}, S_i^{(j+1)}\} = \text{Aggregator}(f_1, f_2, f_3). 
\end{equation}

In this consensus output, $F_{pos}$ summarizes correctly preserved content and successfully executed modifications to be maintained in subsequent iterations, $F_{neg}$ pinpoints specific remaining defects (e.g., ``wrong object color'') to guide the next correction, and $S_i^{(j+1)} \in [0, 10]$ denotes the average score assigned by the three experts. This score determines whether the current image is accepted as the current sub-task output or triggers a regeneration loop.

\begin{figure}[t]
    \centering
    \begin{minipage}[c]{0.48\linewidth}
        \centering
        \includegraphics[width=\linewidth]{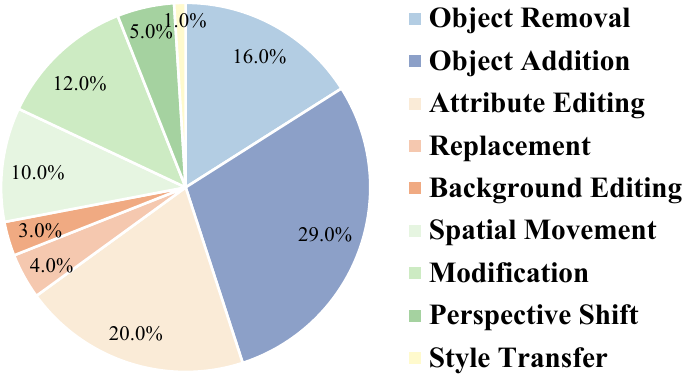}
    \end{minipage}
    \hfill
    \begin{minipage}[c]{0.48\linewidth}
        \centering
        \resizebox{0.65\linewidth}{!}{
            \setlength{\tabcolsep}{3pt}
            \begin{tabular}{c r}
                \toprule
                \multicolumn{2}{c}{\textbf{Turn-Wise Distribution}} \\
                \midrule
                \textbf{Turns} & \textbf{Samples} \\
                \midrule
                4   & 527 \\
                5   & 172 \\
                6   & 216 \\
                7   & 59 \\
                8+   & 26 \\
                \bottomrule
            \end{tabular}
        }
    \end{minipage}
    \caption{\textbf{Statistics of MTEditBench.} The figure displays the distribution of editing operation types and sequence lengths.}
    \label{fig:MTEditBench}
    \vspace{-0.3cm}
\end{figure}

\begin{table*}[t]
\caption{\textbf{Performance comparison on the MTEditBench dataset.}
Bold numbers indicate the best results among the evaluated open-source baselines and IMAGAgent, and underlined numbers indicate the second-best results.
Entries in \textcolor{gray}{gray} denote proprietary models.}
\label{table:onSelfBuilt}
\centering
\scriptsize
\setlength{\tabcolsep}{1.3pt}
\renewcommand{\arraystretch}{1.05}
\resizebox{\linewidth}{!}{%
\begin{tabular}{l|cccc|cccc|cccc|cccc|cccc|cccc}
\toprule
\multirow{2}{*}{\textbf{Method}}
& \multicolumn{4}{c|}{\textbf{Turn-1}}
& \multicolumn{4}{c|}{\textbf{Turn-2}}
& \multicolumn{4}{c|}{\textbf{Turn-3}}
& \multicolumn{4}{c|}{\textbf{Turn-4}}
& \multicolumn{4}{c|}{\textbf{Turn-5}}
& \multicolumn{4}{c}{\textbf{Average}} \\
\cmidrule(lr){2-5}
\cmidrule(lr){6-9}
\cmidrule(lr){10-13}
\cmidrule(lr){14-17}
\cmidrule(lr){18-21}
\cmidrule(lr){22-25}

& DINO$\uparrow$ & CLIP-I$\uparrow$ & CLIP-T$\uparrow$ & AugCLIP$\uparrow$
& DINO$\uparrow$ & CLIP-I$\uparrow$ & CLIP-T$\uparrow$ & AugCLIP$\uparrow$
& DINO$\uparrow$ & CLIP-I$\uparrow$ & CLIP-T$\uparrow$ & AugCLIP$\uparrow$
& DINO$\uparrow$ & CLIP-I$\uparrow$ & CLIP-T$\uparrow$ & AugCLIP$\uparrow$
& DINO$\uparrow$ & CLIP-I$\uparrow$ & CLIP-T$\uparrow$ & AugCLIP$\uparrow$
& DINO$\uparrow$ & CLIP-I$\uparrow$ & CLIP-T$\uparrow$ & AugCLIP$\uparrow$ \\
\midrule

ACE++~\cite{mao2025ace++}
& 0.529 & 0.806 & 0.153 & 0.287
& 0.526 & 0.803 & 0.151 & 0.260
& 0.517 & 0.793 & 0.148 & 0.236
& 0.501 & 0.777 & 0.142 & 0.211
& 0.484 & 0.761 & 0.136 & 0.188
& 0.511 & 0.788 & 0.146 & 0.236 \\

HQEdit~\cite{hui2024hq}
& 0.677 & 0.813 & 0.236 & 0.276
& 0.674 & 0.802 & 0.235 & 0.248
& 0.669 & 0.809 & 0.238 & 0.224
& 0.662 & 0.799 & 0.235 & 0.200
& 0.650 & 0.793 & 0.234 & 0.174
& 0.666 & 0.803 & 0.236 & 0.224 \\

UltraEdit~\cite{zhao2024ultraedit}
& 0.762 & 0.869 & \underline{0.251} & 0.293
& 0.758 & 0.865 & \underline{0.247} & 0.269
& 0.746 & \underline{0.855} & 0.243 & 0.245
& 0.722 & 0.837 & 0.233 & 0.222
& 0.698 & 0.820 & 0.223 & 0.199
& 0.737 & 0.849 & 0.239 & 0.246 \\

ICEdit~\cite{zhang2025context}
& 0.763 & 0.847 & 0.250 & 0.302
& 0.759 & 0.843 & 0.246 & 0.280
& 0.747 & 0.833 & 0.243 & 0.259
& 0.723 & 0.816 & 0.233 & 0.237
& \underline{0.699} & 0.799 & 0.223 & 0.215
& 0.738 & 0.828 & 0.239 & 0.259 \\

VAREdit~\cite{mao2025visual}
& 0.766 & 0.863 & 0.249 & 0.298
& 0.762 & 0.859 & 0.244 & 0.274
& \underline{0.750} & 0.849 & 0.242 & 0.250
& \underline{0.726} & 0.831 & 0.235 & 0.228
& 0.698 & 0.814 & 0.228 & 0.206
& \underline{0.740} & 0.843 & 0.240 & 0.251 \\

VINCIE~\cite{qu2025vincie}
& 0.787 & 0.889 & 0.246 & \underline{0.309}
& \underline{0.767} & 0.873 & 0.245 & \underline{0.290}
& 0.736 & 0.853 & \underline{0.244} & \underline{0.271}
& 0.708 & \underline{0.848} & \underline{0.243} & \underline{0.252}
& 0.696 & \underline{0.837} & \underline{0.240} & \underline{0.233}
& 0.739 & \underline{0.860} & \underline{0.244} & \underline{0.271} \\

OmniGen~\cite{xiao2025omnigen}
& \underline{0.800} & \underline{0.896} & 0.247 & 0.306
& 0.766 & \underline{0.874} & 0.242 & 0.284
& 0.720 & 0.848 & 0.242 & 0.263
& 0.588 & 0.783 & 0.239 & 0.242
& 0.480 & 0.723 & 0.236 & 0.221
& 0.671 & 0.825 & 0.241 & 0.263 \\

\textcolor{gray}{GPT-4o}~\cite{hurst2024gpt}
& \textcolor{gray}{0.776}
& \textcolor{gray}{0.878}
& \textcolor{gray}{0.262}
& \textcolor{gray}{0.314}
& \textcolor{gray}{0.763}
& \textcolor{gray}{0.862}
& \textcolor{gray}{0.257}
& \textcolor{gray}{0.298}
& \textcolor{gray}{0.742}
& \textcolor{gray}{0.849}
& \textcolor{gray}{0.248}
& \textcolor{gray}{0.282}
& \textcolor{gray}{0.698}
& \textcolor{gray}{0.817}
& \textcolor{gray}{0.241}
& \textcolor{gray}{0.267}
& \textcolor{gray}{0.657}
& \textcolor{gray}{0.786}
& \textcolor{gray}{0.235}
& \textcolor{gray}{0.252}
& \textcolor{gray}{0.727}
& \textcolor{gray}{0.838}
& \textcolor{gray}{0.249}
& \textcolor{gray}{0.283} \\

\midrule

\textbf{IMAGAgent}
& \textbf{0.803} & \textbf{0.897} & \textbf{0.254} & \textbf{0.319}
& \textbf{0.790} & \textbf{0.890} & \textbf{0.250} & \textbf{0.307}
& \textbf{0.769} & \textbf{0.871} & \textbf{0.247} & \textbf{0.298}
& \textbf{0.745} & \textbf{0.861} & \textbf{0.246} & \textbf{0.289}
& \textbf{0.721} & \textbf{0.854} & \textbf{0.243} & \textbf{0.282}
& \textbf{0.766} & \textbf{0.875} & \textbf{0.248} & \textbf{0.299} \\

\bottomrule
\end{tabular}%
}
\vspace{-0.2cm}
\end{table*}

\noindent\textbf{Iterative Refinement Strategy.}
The feedback loop is governed by an acceptance threshold and a maximum iteration budget. Upon deriving the consensus feedback, we update the context $\mathcal{C} \leftarrow \mathcal{C} \cup \{\mathcal{R}_i^{(j)}, \mathcal{I}_i^{(j+1)}, \mathcal{F}_i^{(j+1)}\}$ to append the current edit attempt to the historical context. Subsequently, the process branches based on the quality score $S_i^{(j+1)}$.
If $S_i^{(j+1)} \ge \tau_{sr}$, the current image is accepted as the current sub-task output. Conversely, if the score falls below the threshold and the iteration limit is not reached, the system triggers a regeneration step: the iteration counter increments ($j \leftarrow j + 1$), and the agent leverages the negative feedback $F_{neg}$ within the updated $\mathcal{C}$ to rectify errors in the next iteration.
To bound the refinement process, each sub-task is allowed at most
$\tau_{it}$ editing attempts. If no candidate meets the acceptance
threshold within this budget, the system selects the candidate with
the highest recorded consensus score as a best-effort output:
\begin{equation}
    \mathcal{I}_i =
    \begin{cases}
      \mathcal{I}_i^{(j+1)}
      & \text{if } S_i^{(j+1)} \ge \tau_{sr}, \\[2pt]
      \mathcal{I}_i^{(k^\star)}
      & \text{if } S_i^{(j+1)} < \tau_{sr}
        \text{ and } j+1 = \tau_{it},
    \end{cases}
\end{equation}
where $k^\star = \arg\max_{1 \le k \le \tau_{it}} S_i^{(k)}$
indexes the highest-scoring candidate among the completed attempts.
The fallback output may remain below the acceptance threshold.
The selected output is passed to the next sub-task, while historical
plans and critique feedback are retained in $\mathcal{C}$ to inform
tool-chain orchestration.

\section{Experiments}~\label{sec:exp}
\subsection{Datasets}

\noindent\textbf{MagicBrush.}
For evaluation, we utilize MagicBrush~\cite{zhang2023magicbrush}, a manually annotated dataset containing over 10k triplets sourced from 5,313 sessions. However, its limited depth (maximum three turns) constrains the assessment of structural consistency and semantic stability in long-horizon editing workflows.

\noindent\textbf{MTEditBench.}
To evaluate stability in prolonged interactions, we introduce MTEditBench, which comprises 1,000 high-quality sequences with depths of at least four turns, specifically curated to assess cumulative semantic consistency in complex multi-turn editing scenarios. We construct MTEditBench from the large-scale VINCIE training corpus~\cite{qu2025vincie} through a structured screening and verification protocol.
The construction protocol consists of four stages. First, we perform heuristic segmentation on raw editing instructions using punctuation and conjunction cues to identify samples with potential multi-turn structure; candidate samples are then randomly shuffled to mitigate ordering and distribution bias.
Second, we process the candidates with an image-conditioned VLM validator (GLM-4.1V-9B-Thinking~\cite{hong2025glm}) to perform turn segmentation and semantic validity checking for each proposed editing step.
Third, to guarantee long-horizon interaction depth, we retain only candidates in which the VLM identifies at least four valid editing turns; during this stage, computationally intensive reasoning modes are disabled to balance large-scale processing efficiency with validation reliability.
Fourth, we conduct manual verification to remove low-quality or semantically inconsistent samples while keeping the dataset curation pipeline separate from the generative and reflective models used in our agent.
Specifically, we check whether: (i) each turn is executable and semantically coherent; (ii) the instruction chain avoids contradictory downstream edits; and (iii) the image quality supports stable evaluation. By explicitly separating the dataset curation pipeline (GLM-4.1V and human reviewers) from our agent's generative and reflective models (Qwen and Doubao), we reduce potential bias from overlap with the generative and reflective models.
After filtering and verification, MTEditBench contains 1,000 long-horizon interaction sequences covering diverse operations, including object removal, object addition, and attribute editing. Fig.~\ref{fig:MTEditBench} presents the detailed dataset statistics, including operation-type distribution and sequence-length distribution across interaction turns. In the benchmark evaluation, Turn-$t$ refers to the $t$-th instruction in the original interaction sequence.

\subsection{Metrics and Implementation Details}

\noindent\textbf{Metrics.}
To comprehensively evaluate multi-turn image editing, we adopt DINO~\cite{caron2021emerging}, CLIP-I, and CLIP-T~\cite{shafiullah2022clip,hessel2021clipscore}, following prior work to measure visual consistency, image similarity, and text--image alignment, respectively. On MTEditBench, we additionally report AugCLIP~\cite{kim2025preserve}, a recently introduced metric tailored to text-guided image editing. AugCLIP adaptively balances source-image preservation and target-text modification in the CLIP feature space, providing a more context-aware assessment of editing quality.

\noindent\textbf{Implementations.}
All experiments are conducted on a single NVIDIA A100 GPU. Proprietary models are invoked through APIs, while the remaining components are deployed locally.
For the planning stage, we adopt Qwen-VL-MAX as the constraint-aware planner to decompose global instructions into executable atomic sub-tasks while preserving long-sequence structural integrity.
For tool-chain orchestration, we use GLM-4.1V-9B-Thinking as the editing controller to dynamically schedule heterogeneous tools according to the current image state and sub-task requirements.
The invocable tool set includes SAM~\cite{kirillov2023segment} for segmentation, Grounding-DINO-Base for open-set detection, CLIP-ViT-Large-Patch14 for image retrieval, and three editing backbones (Doubao-Seedream-4.0, Qwen-Image-Edit, and Stable-Diffusion-XL-Base-1.0) for generation and local refinement.
For reflection, we employ a multi-expert committee consisting of Qwen-VL-MAX, Doubao-Seed-1.6-Vision, and Doubao-Seed-1.6; their critiques are aggregated by DeepSeek-V3.2 to produce final consensus feedback. In the single-expert ablation, Qwen-VL-MAX is used as the sole critic.
To balance quality and efficiency in iterative editing, we set the maximum refinement iterations to $\tau_{it}=3$ and the acceptance threshold to $\tau_{sr}=7$.
For the backbone-controlled analysis, we additionally construct \textit{IMAGAgent (GPT-4o)}, which fixes GPT-4o as the sole editing backbone while retaining the framework components and inference settings. This controlled variant is used only in TABLE~\ref{tab:backbone_controlled} and is distinct from the default dynamically orchestrated configuration used in the main comparison.

\begin{table*}[t]
\caption{\textbf{Performance comparison on the MagicBrush benchmark.} Bold numbers indicate the best results among the evaluated open-source baselines and IMAGAgent, and underlined numbers indicate the second-best results. Entries in \textcolor{gray}{gray} denote proprietary models.}
\label{table:onMagicBrush}
\centering
\small
\setlength{\tabcolsep}{3pt}
\resizebox{1.0\linewidth}{!} {
\begin{tabular}{l|ccc|ccc|ccc|ccc}
\toprule
\multirow{2}{*}{\textbf{Method}} 
& \multicolumn{3}{c|}{\textbf{Turn-1}} 
& \multicolumn{3}{c|}{\textbf{Turn-2}} 
& \multicolumn{3}{c|}{\textbf{Turn-3}} 
& \multicolumn{3}{c}{\textbf{Average}} \\
\cmidrule(lr){2-4} \cmidrule(lr){5-7} \cmidrule(lr){8-10} \cmidrule(lr){11-13}
& DINO$\uparrow$ & CLIP-I$\uparrow$ & CLIP-T$\uparrow$ 
& DINO$\uparrow$ & CLIP-I$\uparrow$ & CLIP-T$\uparrow$ 
& DINO$\uparrow$ & CLIP-I$\uparrow$ & CLIP-T$\uparrow$
& DINO$\uparrow$ & CLIP-I$\uparrow$ & CLIP-T$\uparrow$ \\
\midrule
ACE++~\cite{mao2025ace++}
& 0.522 & 0.791 & 0.176
& 0.491 & 0.770 & 0.169
& 0.475 & 0.748 & 0.163
& 0.496 & 0.770 & 0.169 \\
HQEdit~\cite{hui2024hq}
& 0.522 & 0.696 & 0.259 
& 0.441 & 0.659 & 0.248 
& 0.397 & 0.637 & 0.238
& 0.453 & 0.664 & 0.248 \\
UltraEdit~\cite{zhao2024ultraedit}
& 0.755 & 0.852 & \textbf{0.289}
& 0.706 & 0.827 & \underline{0.278} 
& 0.683 & \underline{0.810} & 0.266
& 0.715 & 0.830 & \underline{0.278} \\
ICEdit~\cite{zhang2025context}
& 0.757 & 0.833 & 0.281
& 0.708 & 0.806 & 0.277
& 0.681 & 0.786 & 0.265
& 0.715 & 0.808 & 0.274 \\
VAREdit~\cite{mao2025visual}
& 0.757 & 0.847 & 0.276
& 0.712 & 0.825 & 0.264
& \underline{0.689} & 0.806 & 0.258
& 0.719 & 0.826 & 0.266 \\
VINCIE~\cite{qu2025vincie}
& 0.838 & 0.906 & 0.272
& \underline{0.721} & 0.848 & 0.272
& 0.645 & 0.804 & \underline{0.271}
& \underline{0.735} & 0.853 & 0.272 \\
OmniGen~\cite{xiao2025omnigen}
& \textbf{0.874} & \underline{0.924} & 0.273 
& 0.718 & \underline{0.851} & 0.264
& 0.586 & 0.786 & 0.261
& 0.726 & \underline{0.854} & 0.266 \\
\textcolor{gray}{GPT-4o}~\cite{hurst2024gpt}
& \textcolor{gray}{0.805} & \textcolor{gray}{0.875} & \textcolor{gray}{0.293} 
& \textcolor{gray}{0.708} & \textcolor{gray}{0.820} & \textcolor{gray}{0.300} 
& \textcolor{gray}{0.666} & \textcolor{gray}{0.789} & \textcolor{gray}{0.292}
& \textcolor{gray}{0.726} & \textcolor{gray}{0.828} & \textcolor{gray}{0.295} \\
\midrule
\textbf{IMAGAgent}                 
& \underline{0.862} & \textbf{0.929} & \underline{0.282}
& \textbf{0.791} & \textbf{0.893} & \textbf{0.284}
& \textbf{0.766} & \textbf{0.872} & \textbf{0.279}
& \textbf{0.806} & \textbf{0.898} & \textbf{0.282} \\
\bottomrule
\end{tabular}
}
   \vspace{-0.3cm}
\end{table*}

\subsection{Main Results}

We compare IMAGAgent with representative state-of-the-art image editing methods, including ACE++~\cite{mao2025ace++}, HQEdit~\cite{hui2024hq}, UltraEdit~\cite{zhao2024ultraedit}, ICEdit~\cite{zhang2025context}, VAREdit~\cite{mao2025visual}, VINCIE~\cite{qu2025vincie}, OmniGen~\cite{xiao2025omnigen}, and GPT-4o~\cite{hurst2024gpt}. We evaluate the complete systems on MTEditBench and MagicBrush, followed by a controlled comparison under a fixed editing backbone.

\noindent\textbf{Quantitative Evaluation.}
TABLE~\ref{table:onSelfBuilt} reports the quantitative results on MTEditBench. IMAGAgent achieves the best results among publicly available methods across the four metrics and exhibits less degradation as the editing sequence becomes longer. For example, OmniGen's~\cite{xiao2025omnigen} DINO score decreases from 0.800 at Turn-1 to 0.480 at Turn-5, whereas IMAGAgent maintains a score of 0.721 at Turn-5. IMAGAgent also achieves the highest AugCLIP score at every turn, with an average of 0.299, compared with 0.271 for the strongest public baseline VINCIE and 0.283 for GPT-4o. These results indicate stronger preservation of visual and semantic consistency over extended editing trajectories.
TABLE~\ref{table:onMagicBrush} presents the results on MagicBrush~\cite{zhang2023magicbrush}. Although methods such as OmniGen~\cite{xiao2025omnigen} and UltraEdit~\cite{zhao2024ultraedit} remain competitive at Turn-1, their performance decreases more noticeably in subsequent turns. By Turn-3, IMAGAgent outperforms all compared publicly available baselines across DINO, CLIP-I, and CLIP-T, indicating stronger structural preservation, visual identity retention, and instruction alignment. Together, the two benchmarks provide complementary evaluations: MagicBrush assesses generalization in short-horizon interactions, whereas MTEditBench emphasizes stability under longer and more strongly coupled editing sequences.

\begin{figure*}[t]
    \centering
    \includegraphics[width=0.9\linewidth]{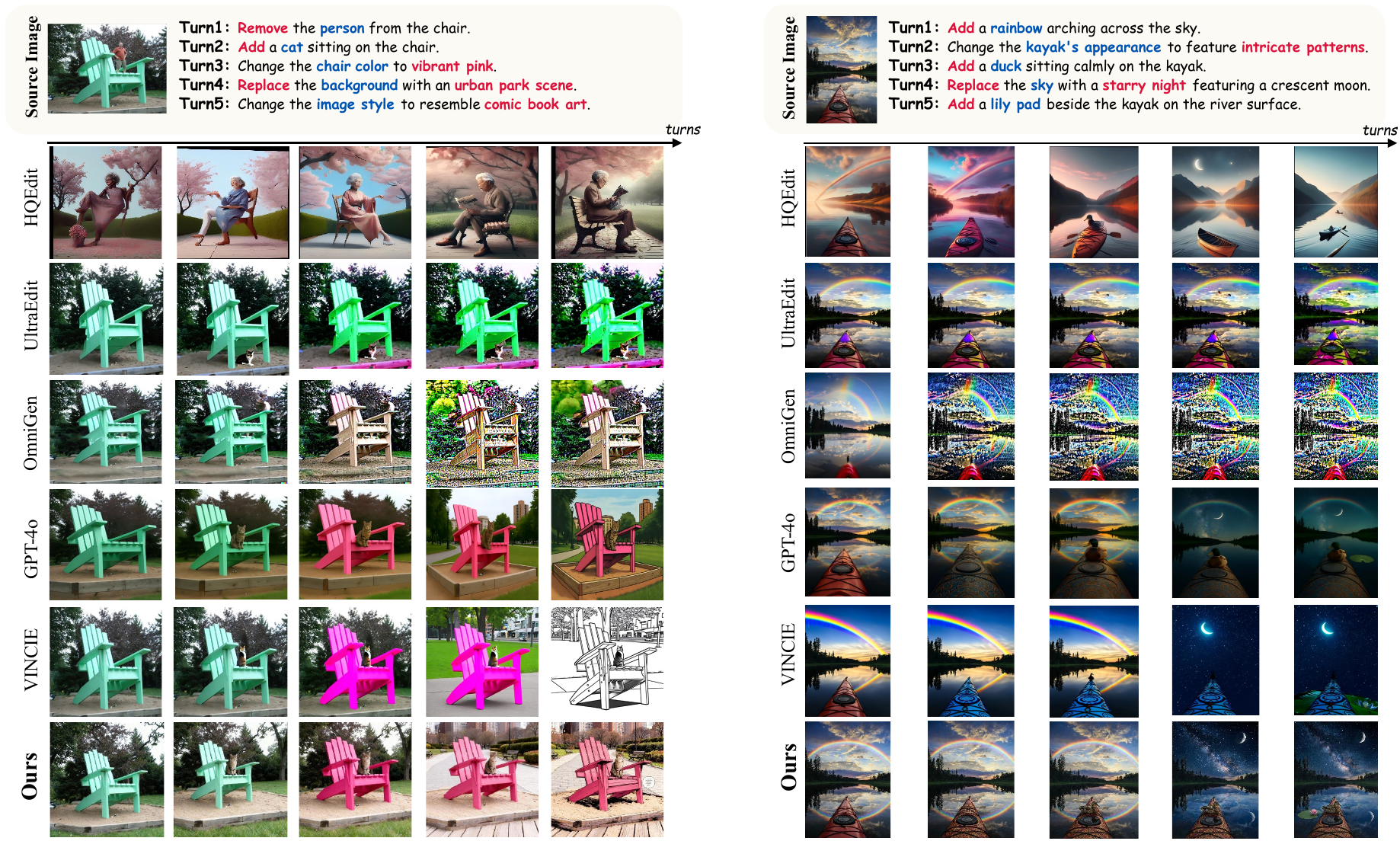}
       \vspace{-0.1cm}
    \caption{\textbf{Qualitative results on MTEditBench.} Comparisons show IMAGAgent achieves top visual quality.}
    \label{fig:compare}
    \vspace{-0.3cm}
\end{figure*}

\begin{figure}[t]
    \centering
    \includegraphics[width=0.95\linewidth]{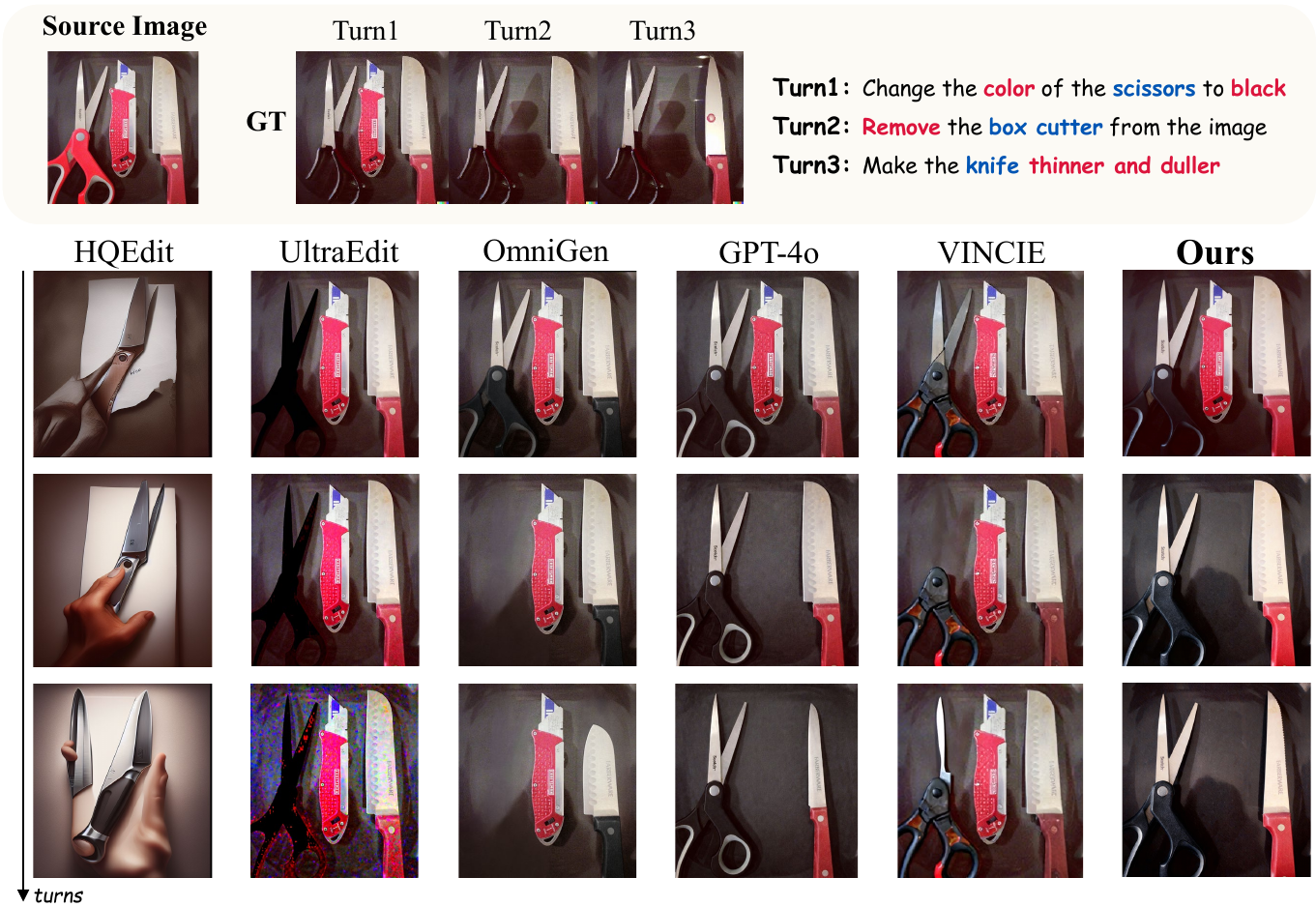}
       \vspace{-0.2cm}
    \caption{\textbf{Qualitative results on MagicBrush.} IMAGAgent outperforms SOTA in complex spatial and attribute constraints.}
    \label{fig:compare2}
    \vspace{-0.4cm}
\end{figure}

\begin{table}[t]
\centering
\caption{\textbf{Backbone-controlled comparison on MTEditBench.}
Standalone editors directly process the original turn-level instruction sequence, using each preceding output as the input to the next turn.  Best results are in \textbf{bold}, and second-best results are \underline{underlined}.}
\label{tab:backbone_controlled}
\small
\setlength{\tabcolsep}{4.2pt}
\renewcommand{\arraystretch}{1.08}
\resizebox{\linewidth}{!}{%
\begin{tabular}{lccccccccc}
\toprule
\multirow{2}{*}{\textbf{Method}}
& \multicolumn{3}{c}{\textbf{Turn-3}}
& \multicolumn{3}{c}{\textbf{Turn-4}}
& \multicolumn{3}{c}{\textbf{Turn-5}} \\
\cmidrule(lr){2-4}
\cmidrule(lr){5-7}
\cmidrule(lr){8-10}
& \textbf{DINO$\uparrow$}
& \textbf{CLIP-I$\uparrow$}
& \textbf{CLIP-T$\uparrow$}
& \textbf{DINO$\uparrow$}
& \textbf{CLIP-I$\uparrow$}
& \textbf{CLIP-T$\uparrow$}
& \textbf{DINO$\uparrow$}
& \textbf{CLIP-I$\uparrow$}
& \textbf{CLIP-T$\uparrow$} \\
\midrule

Doubao-Seedream-4.0 (Standalone)
& 0.653 & 0.742 & 0.156
& 0.611 & 0.685 & 0.121
& 0.548 & 0.596 & 0.107 \\

Qwen-Image-Edit (Standalone)
& 0.668 & 0.731 & 0.149
& 0.598 & 0.694 & 0.128
& 0.561 & 0.583 & 0.099 \\

\midrule

GPT-4o (Standalone)
& \underline{0.742}
& \underline{0.849}
& \underline{0.248}
& \underline{0.698}
& \underline{0.817}
& \underline{0.241}
& \underline{0.657}
& \underline{0.786}
& \underline{0.235} \\

\textbf{IMAGAgent (GPT-4o)}
& \textbf{0.791}
& \textbf{0.886}
& \textbf{0.258}
& \textbf{0.772}
& \textbf{0.875}
& \textbf{0.255}
& \textbf{0.753}
& \textbf{0.866}
& \textbf{0.252} \\

\bottomrule
\end{tabular}%
}
\vspace{-0.3cm}
\end{table}

\noindent\textbf{Backbone-Controlled Comparison.}
To isolate framework-level improvements from the capacity of the underlying editing backbone, we conduct a matched backbone-controlled comparison on MTEditBench. Specifically, we fix GPT-4o as the editing backbone and evaluate two configurations on the same samples and original turn-level instruction sequences. The standalone GPT-4o editor processes each instruction sequentially using the output of the preceding turn, without constraint-aware planning, explicit trajectory-level context, auxiliary tool coordination, or feedback-driven refinement. In contrast, IMAGAgent (GPT-4o) retains the same editing backbone while incorporating constraint-aware planning, history-conditioned tool-chain orchestration, and multi-expert collaborative reflection. We additionally include standalone Doubao-Seedream-4.0 and Qwen-Image-Edit as backbone references.
As shown in TABLE~\ref{tab:backbone_controlled}, all standalone editors exhibit increasingly pronounced degradation as the editing sequence becomes longer. More importantly, under the matched GPT-4o backbone, IMAGAgent consistently improves all three metrics from Turn-3 to Turn-5. The DINO gain increases from 0.049 at Turn-3 to 0.096 at Turn-5, while the Turn-5 CLIP-I and CLIP-T scores improve by 0.080 and 0.017, respectively. The widening performance margin across successive turns indicates that the benefit of IMAGAgent becomes increasingly substantial as editing constraints and intermediate errors accumulate. These results demonstrate that the performance gains cannot be attributed solely to the editing backbone, but instead arise from the proposed closed-loop mechanism for maintaining cross-turn constraints and correcting intermediate failures.

\noindent\textbf{Qualitative Evaluation.} Fig.~\ref{fig:compare} and Fig.~\ref{fig:compare2} provide qualitative comparisons of IMAGAgent against state-of-the-art (SOTA) approaches on the MTEditBench and MagicBrush datasets. As illustrated in Fig.~\ref{fig:compare}, baseline models suffer from severe error accumulation and semantic drift during long-horizon editing. At Turn-4 and Turn-5, OmniGen introduces heavy visual artifacts, while HQEdit~\cite{hui2024hq} and VINCIE~\cite{qu2025vincie} struggle to maintain the original subject's identity or hallucinate unintended stylistic shifts. In contrast, IMAGAgent effectively executes complex sequential instructions, such as adding entities, altering backgrounds, and applying global styles, while preserving the structural integrity of unedited regions.
Furthermore, Fig.~\ref{fig:compare2} highlights IMAGAgent's superiority in high-constraint spatial and attribute manipulations. When tasked with removing the box cutter and geometrically modifying the knife (i.e., making it thinner and duller), several baselines fail in this setting: UltraEdit~\cite{zhao2024ultraedit} suffers from complete image degradation by Turn-3, HQEdit~\cite{hui2024hq} hallucinates unrelated elements like human hands, and others fail to preserve the background. Benefiting from constraint-aware planning and multi-expert collaborative reflection, IMAGAgent executes these precise local edits with limited semantic drift. Overall, our approach delivers the strongest qualitative results, with better instruction adherence and long-term stability across multi-turn workflows.

\begin{figure}
        \centering
    \includegraphics[width=0.98\linewidth]{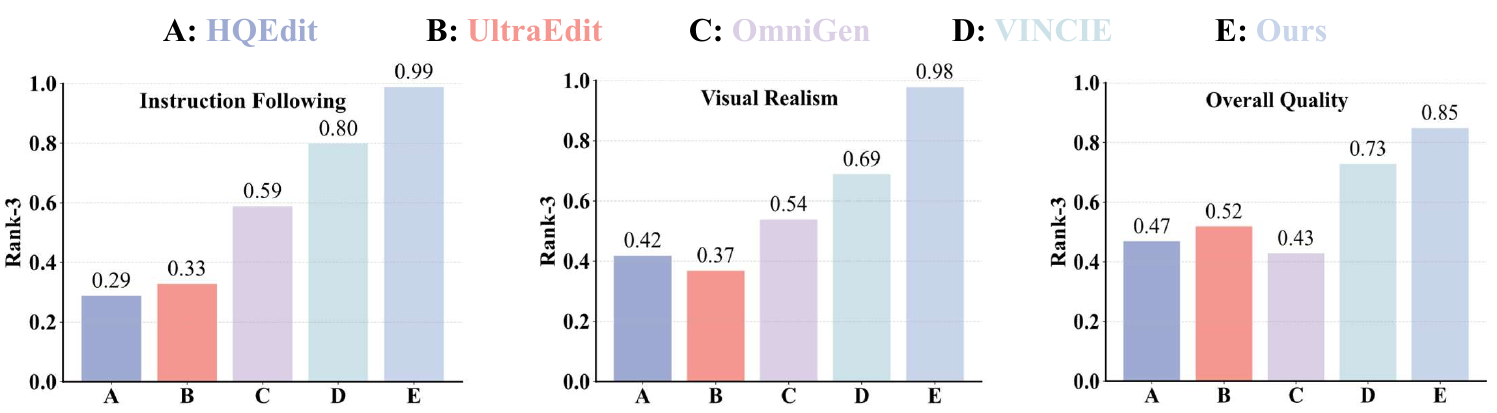}
       \vspace{-0.2cm}
    \caption{\textbf{User study} examined from three perspectives: instruction following, visual realism, and overall quality.}
    \label{fig:user_study}
    \vspace{-0.4cm}
\end{figure}

\noindent\textbf{User Study.} We randomly selected 50 editing cases, shuffled the outputs from various methods, and recruited 50 participants (25 male, 25 female) to select their top three preferred results separately for instruction following, visual realism, and overall editing quality, with Rank-3 denoting the proportion of times each method was selected among the top three. 
From Fig.~\ref{fig:user_study}, IMAGAgent receives the highest overall preference among the compared methods, particularly in instruction adherence and visual realism. These results suggest that the algorithmic robustness of IMAGAgent translates into improved human-perceived visual quality and more stable multi-turn image editing.

\subsection{Ablation Study}
\begin{figure*}[t]
    \centering
    \includegraphics[width=0.95\linewidth]{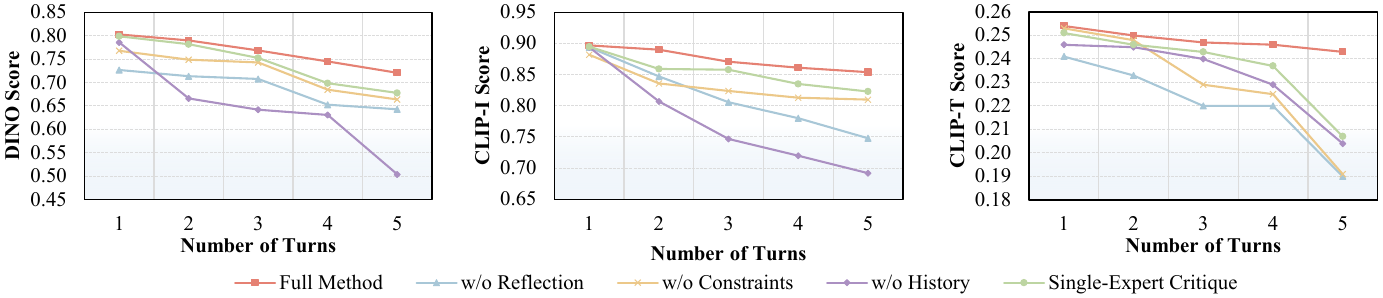}
    \vspace{-0.2cm}
    \caption{\textbf{Turn-wise diagnostic ablation on MTEditBench.} The plots show how representative mechanism-level variants degrade across successive editing turns.}
    \label{fig:ablation}
    \vspace{-0.3cm}
\end{figure*}

To isolate the contribution of each agentic component, we conduct controlled ablation experiments under the same evaluation protocol, where each variant removes or replaces only the indicated component. TABLE~\ref{tab:ablation} reports five component-level variants, while Fig.~\ref{fig:ablation} provides complementary turn-wise diagnostic results to characterize degradation trends across successive editing turns. The \textit{w/o C-Plan} variant replaces constraint-aware planning with naive sentence-level decomposition, without dependency-based reordering, redundant-operation consolidation, or visual-feasibility filtering. The \textit{w/o Tool Orch.} variant removes dynamic tool-chain orchestration and applies a fixed editing model to all sub-tasks. The \textit{w/o History} variant discards the accumulated editing trajectory, while \textit{w/o Reflection} disables visual evaluation and feedback-driven retries. Finally, \textit{Single-Expert} retains the refinement loop but replaces the heterogeneous critic panel with a single VLM. In Fig.~\ref{fig:ablation}, \textit{w/o Constraints} removes explicit preservation of previously established editing constraints.

\noindent\textbf{Effect of Constraint-Aware Planning and Tool Orchestration.}
Replacing constraint-aware planning with naive decomposition reduces the five-turn average DINO score from 0.766 to 0.757, showing that structured decomposition, dependency-aware reordering, and feasibility filtering provide a consistent improvement over simple sentence-level splitting. Removing dynamic tool orchestration causes a larger degradation, reducing DINO from 0.766 to 0.727 and CLIP-I from 0.875 to 0.835. This result indicates that a fixed editor is insufficient for heterogeneous editing operations, which may require different combinations of detection, segmentation, retrieval, local editing, and global generation.

\begin{table}[t]
\centering
\caption{\textbf{Component ablation on MTEditBench.}
All variants follow the same evaluation protocol and retain the full configuration except for the indicated component. Best results are in \textbf{bold}, and second-best results are \underline{underlined}.}
\vspace{-0.2cm}
\label{tab:ablation}
\resizebox{\linewidth}{!}{
\begin{tabular}{lcccccccc}
\toprule
\multirow{2}{*}{\textbf{Variant}}
& \multicolumn{5}{c}{\textbf{Components}}
& \multicolumn{3}{c}{\textbf{Metrics}} \\
\cmidrule(lr){2-6}
\cmidrule(lr){7-9}
& \textbf{C-Plan}
& \textbf{Tool}
& \textbf{Hist.}
& \textbf{Refl.}
& \textbf{Multi.}
& \textbf{DINO$\uparrow$}
& \textbf{CLIP-I$\uparrow$}
& \textbf{CLIP-T$\uparrow$} \\
\midrule

w/o C-Plan
& \xmark & \cmark & \cmark & \cmark & \cmark
& \underline{0.757}
& \underline{0.869}
& \underline{0.247} \\

w/o Tool Orch.
& \cmark & \xmark & \cmark & \cmark & \cmark
& 0.727
& 0.835
& 0.238 \\

w/o History
& \cmark & \cmark & \xmark & \cmark & \cmark
& 0.646
& 0.772
& 0.233 \\

w/o Reflection
& \cmark & \cmark & \cmark & \xmark & --
& 0.689
& 0.815
& 0.221 \\

Single-Expert
& \cmark & \cmark & \cmark & \cmark & \xmark
& 0.742
& 0.854
& 0.237 \\

\midrule

\textbf{IMAGAgent}
& \cmark & \cmark & \cmark & \cmark & \cmark
& \textbf{0.766}
& \textbf{0.875}
& \textbf{0.248} \\

\bottomrule
\end{tabular}
}
\vspace{-0.2cm}
\end{table}

\noindent\textbf{Importance of Historical Context and Reflection.}
Historical context is critical for preserving previously established visual constraints. Removing the editing history produces the largest degradation in DINO and CLIP-I, which decrease to 0.646 and 0.772, respectively. Without access to prior edits and intermediate states, the agent is more likely to overwrite previously modified regions or lose subject identity across turns. Reflection is similarly important for instruction adherence: disabling the feedback loop reduces CLIP-T from 0.248 to 0.221. This confirms that one-pass execution cannot reliably detect and correct intermediate failures before they propagate to subsequent turns.

\noindent\textbf{Benefit of Multi-Expert Reflection.}
Replacing the heterogeneous critic panel with a single VLM results in consistent performance decreases across all three metrics. Although the single-expert variant remains competitive, the full multi-expert design improves DINO from 0.742 to 0.766 and CLIP-I from 0.854 to 0.875. These results suggest that aggregating complementary visual assessments produces more reliable and actionable feedback, reducing the likelihood that structural, semantic, or perceptual defects remain undetected during iterative refinement.

\begin{figure}[t]
    \centering
\includegraphics[width=0.95\linewidth]{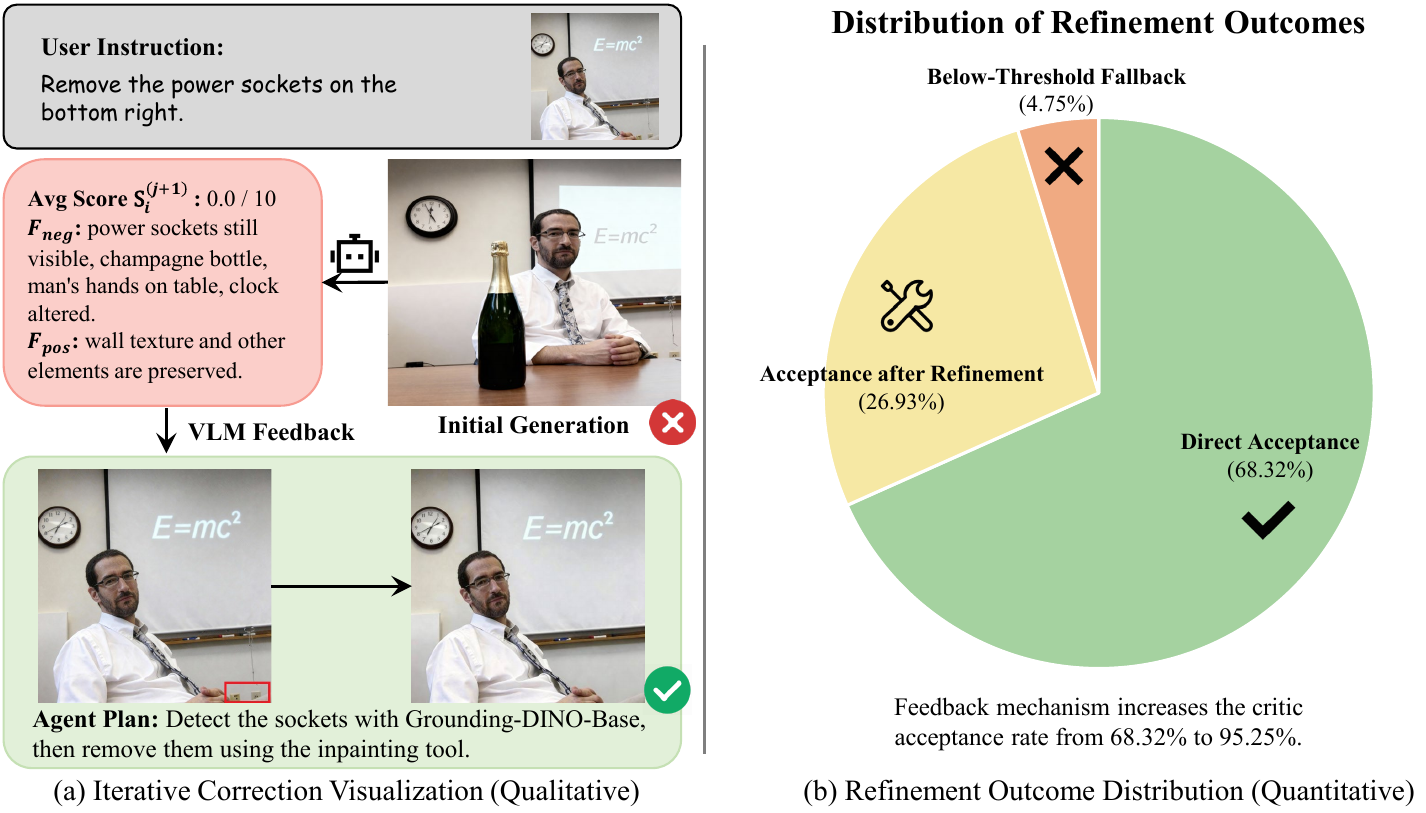}
\caption{\textbf{Feedback-guided refinement.}
(a) An example of targeted correction guided by aggregated expert feedback.
(b) Distribution of direct acceptance, acceptance after refinement,
and below-threshold fallback.
Acceptance is determined by the consensus-score threshold.}
    \label{fig:feedback_analysis}
    \vspace{-0.3cm}
\end{figure}

\subsection{Analysis of Feedback Mechanism}
\label{subsec:feedback_analysis}
\noindent\textbf{Qualitative Correction Capability.}
Fig.~\ref{fig:feedback_analysis} (a) visualizes the feedback-driven correction process. In the "remove power sockets" task, the initial execution failed, causing structural distortion and introducing an unintended bottle. The reflection module identified these defects as negative signals while specifying structural preservation as a positive constraint. 
Guided by this actionable critique, the agent optimized tool selection, using Grounding-DINO-Base for detection and inpainting for removal, to obtain the desired result. 
This demonstrates that multi-expert reflection provides actionable, fine-grained feedback that corrects structural failures and stabilizes multi-turn editing beyond simple binary judgment.

\noindent\textbf{Refinement Efficacy.}
We statistically analyze the refinement trajectory of IMAGAgent on the MagicBrush validation set. As shown in Fig.~\ref{fig:feedback_analysis} (b), 68.32\% of the editing steps meet the acceptance threshold ($\tau_{sr}$) on the first attempt, while about 27\% are initially rejected but subsequently meet the threshold within the allowed iteration limit ($\tau_{it}=3$). These cases typically involve subtle attribute errors or spatial misalignments that are difficult to handle in a single pass. This result shows that iterative refinement increases the proportion of editing steps accepted within the refinement budget.

\begin{table*}[t]
\centering
\caption{\textbf{Comparison of different feedback policies under the same tool pool.}
All variants share the same planning module, tool set, editing-backbone pool, and maximum retry budget, while differing only in how feedback is generated and utilized.
This comparison assesses the contribution of structured
reflective feedback under a shared maximum editing-attempt budget.
Best results are in \textbf{bold} and the second-best are \underline{underlined}; the entire IMAGAgent row is boldfaced for emphasis.}
\vspace{-0.2cm}
\label{tab:feedback_policy}
\resizebox{0.9\textwidth}{!}{
\begin{tabular}{lcccccccc}
\toprule
\textbf{Variant}
& \textbf{Score}
& \textbf{Text Critique}
& \textbf{Multi-Expert}
& \textbf{Retry}
& \textbf{Avg. Calls$\downarrow$}
& \textbf{DINO$\uparrow$}
& \textbf{CLIP-I$\uparrow$}
& \textbf{CLIP-T$\uparrow$} \\
\midrule

\textcolor{gray}{No Feedback}
& \textcolor{gray}{\xmark}
& \textcolor{gray}{\xmark}
& \textcolor{gray}{\xmark}
& \textcolor{gray}{\xmark}
& \textcolor{gray}{\textbf{1.00}}
& \textcolor{gray}{0.689}
& \textcolor{gray}{0.815}
& \textcolor{gray}{0.221} \\

Score-Only Retry
& \cmark & \xmark & \xmark & \cmark
& 1.75
& 0.705 & 0.828 & 0.226 \\

Best-of-3 Retry
& \cmark & \xmark & \xmark & \cmark
& 3.00
& 0.720 & 0.838 & 0.231 \\

Single-Expert
& \cmark & \cmark & \xmark & \cmark
& \underline{1.42}
& 0.742 & 0.854 & 0.237 \\

Multi w/o Aggregation
& \cmark & \cmark & \cmark & \cmark
& 1.48
& \underline{0.750}
& \underline{0.862}
& \underline{0.241} \\

\midrule
\textbf{IMAGAgent}
& \textbf{\cmark}
& \textbf{\cmark}
& \textbf{\cmark}
& \textbf{\cmark}
& \textbf{{1.40}}
& \textbf{0.766}
& \textbf{0.875}
& \textbf{0.248} \\
\bottomrule
\end{tabular}
}
\vspace{-0.3cm}
\end{table*}

\noindent\textbf{Impact of Different Feedback Policies.}
To examine whether the performance gains primarily arise from structured feedback rather than increased inference computation, we compare different feedback policies in TABLE~\ref{tab:feedback_policy}.
The \textit{Score-Only Retry} and \textit{Best-of-3 Retry} variants require an average of 1.75 and 3.00 calls, respectively, while providing only limited improvements over the no-feedback baseline. This suggests that repeated sampling without explicit textual critiques provides insufficient guidance for correcting specific editing errors.
Introducing textual feedback with the \textit{Single-Expert} variant achieves better performance than \textit{Best-of-3 Retry} while requiring fewer calls (1.42 on average), indicating that explicit and actionable error feedback improves refinement efficiency.
Moreover, \textit{Multi w/o Aggregation} performs consistently below the full IMAGAgent despite using multiple critics. Directly combining raw critiques may introduce redundant or conflicting correction signals during subsequent tool-chain orchestration. In contrast, the LLM aggregator consolidates the multi-expert critiques into unified feedback, resulting in improved DINO, CLIP-I, and CLIP-T scores with an average of 1.40 calls.

\begin{figure*}[t]
    \centering

    \begin{minipage}[c]{0.61\textwidth}
        \centering
        \includegraphics[width=\linewidth]{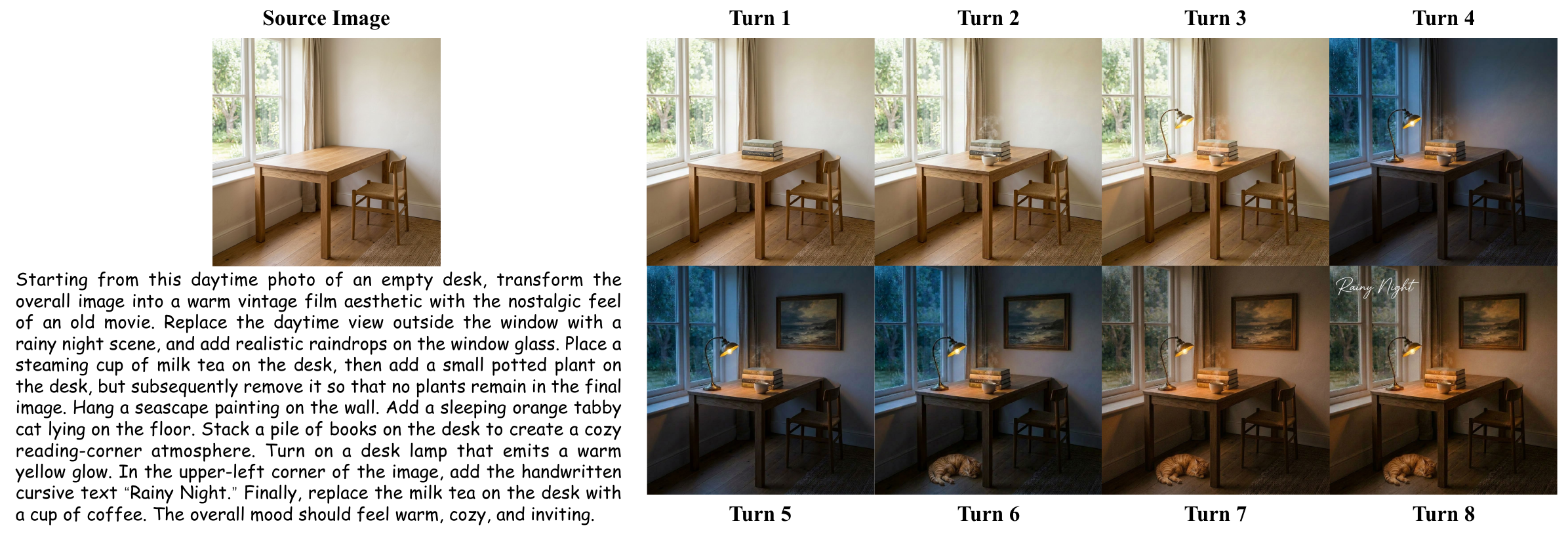}
    \end{minipage}
    \hfill
    \begin{minipage}[c]{0.37\textwidth}
        \centering
        \scriptsize
        \setlength{\tabcolsep}{3pt}
        \renewcommand{\arraystretch}{1.10}

        \resizebox{\linewidth}{!}{%
        \begin{tabular}{c p{0.68\linewidth} c}
            \toprule
            \textbf{No.}
            & \textbf{Edit Operation}
            & \textbf{Planned Order} \\
            \midrule

            1
            & Apply a warm vintage film tone
            & Turn-7 \\

            2
            & Replace the window view with a rainy night scene
              and add raindrops to the glass
            & Turn-4 \\

            3
            & Add a cup of coffee to the desk
            & Turn-2 \\

            4
            & Hang a seascape painting on the wall
            & Turn-5 \\

            5
            & Add a sleeping orange tabby cat lying on the floor
            & Turn-6 \\

            6
            & Place a stack of books on the desk
            & Turn-1 \\

            7
            & Turn on a desk lamp with a warm yellow glow
            & Turn-3 \\

            8
            & Add the handwritten text ``Rainy Night''
              in the upper-left corner
            & Turn-8 \\

            \bottomrule
        \end{tabular}%
        }
    \end{minipage}
     \vspace{-0.2cm}
    \caption{\textbf{Eight-turn editing with full-instruction planning.}
    Given the complete composite instruction, the planner consolidates
    redundant and superseded operations and reorders the retained
    sub-tasks according to their semantic dependencies.
    Each turn corresponds to one planned sub-task.
    Left: the editing trajectory from the source image to Turn-8.
    Right: the execution order produced by the constraint-aware planner.}
    \label{fig:long_horizon_planner}
    \vspace{-0.3cm}
\end{figure*}

\noindent\textbf{Long-Horizon Editing Example.}
Fig.~\ref{fig:long_horizon_planner} presents an eight-turn editing example generated from a single composite instruction. The instruction contains multiple object-level, attribute-level, background, and global-style modifications whose execution order differs from their original textual order. To reduce interference between dependent operations, the constraint-aware planner first performs local content edits, then updates the background and scene elements, and finally applies the global film tone and handwritten text.
The resulting trajectory preserves previously completed edits as new constraints are introduced. For example, the books, coffee, desk lamp, rainy window view, painting, and cat remain consistent through the later style and text modifications. This example illustrates how planner-based reordering and history-conditioned execution support constraint preservation over an extended editing sequence.

\noindent\textbf{Cost vs. Quality Trade-off.}
While the feedback mechanism in the multi-expert reflection section introduces additional inference latency and requires an average of approximately 1.4 editing calls, versus 1.0 for the open-loop baseline, we argue that this trade-off is justified for creative workflows where output quality is paramount.
These additional editing attempts allow the controller to revise tool choices and execution parameters based on specific visual critiques.
Compared with Best-of-3 Retry, IMAGAgent achieves higher DINO, CLIP-I, and CLIP-T scores with fewer average editing calls (1.40 vs. 3.00), supporting the efficiency of targeted refinement.
Please refer to the appendix for further discussion.

\section{Conclusion}~\label{sec:con}
In this work, we proposed IMAGAgent, a robust framework that tackles the error accumulation and semantic drift in multi-turn image editing through a novel ``plan--execute--reflect'' closed-loop mechanism. Unlike traditional open-loop paradigms, IMAGAgent achieves deep synergy among constraint-aware instruction decomposition, dynamic tool-chain orchestration, and multi-expert collaborative reflection. This unified pipeline enables adaptive fine-grained self-correction and adaptive utilization of heterogeneous visual tools. Extensive experiments on MagicBrush and our constructed MTEditBench show that IMAGAgent consistently improves instruction consistency, editing precision, and structural stability over strong baselines, making it a practical framework for long-horizon interactive image editing.

\newpage
\section*{Appendix}

The appendices provide supplementary details on IMAGAgent.
Appendix~\ref{sec:nota} summarizes the notation used throughout the framework.
Appendix~\ref{app:evaluation_protocol} clarifies instruction visibility, planning scope, turn-level evaluation alignment, and controlled comparison settings.
Appendix~\ref{sec:add_qual} presents additional qualitative comparisons on MTEditBench.
Appendix~\ref{sec:discuss} discusses key design choices, refinement settings, and benchmark curation.
Appendix~\ref{sec:cost_analysis} analyzes the API costs of the framework.
Appendix~\ref{sec:prompts} provides the system prompts used in the three core phases of IMAGAgent.
Finally, Appendix~\ref{sec:limitations} discusses the limitations of the framework and directions for future work.

\section{Formal Notation}
\label{sec:nota}

The notation and symbol definitions used in the framework and
evaluation protocol are summarized in TABLE~\ref{notations}.

\begin{table}[t]
    \centering
    \renewcommand{\arraystretch}{1.2}
    \caption{\textbf{Summary of Notation.}}
    \label{notations}
    \resizebox{\linewidth}{!}{%
        \begin{tabular}{c|l}
            \hline
            \textbf{Notation} & \textbf{Definition} \\
            \hline
            $\mathcal{I}_0$
            & Initial source image \\
            $\mathcal{T}_0$
            & Editing instruction provided to the planner \\
            $\mathcal{I}^*$
            & Final edited image \\
            $\mathcal{C}$
            & Historical context containing plans, images, and feedback \\
            $\Phi$
            & Overall mapping function of IMAGAgent \\

            $\mathcal{U}$
            & Original ordered instruction sequence $(u_1,\ldots,u_L)$ \\
            $u_t$
            & Active editing instruction at original Turn-$t$ \\
            $L$
            & Number of original benchmark turns \\

            $\Psi$
            & Constraint-aware planning module \\
            $\mathcal{A}$
            & Ordered sequence of planned sub-tasks $(t_1,\ldots,t_n)$ \\
            $t_i$
            & Instruction for the $i$-th atomic sub-task \\
            $n$
            & Number of sub-tasks after consolidation and reordering \\
            $\mathcal{I}_i$
            & Selected output of sub-task $i$ \\
            $\mathcal{I}_i^{(j)}$
            & Candidate from the $j$-th editing attempt for sub-task $i$ \\
            $\mathcal{R}_i^{(j)}$
            & Tool-chain plan for attempt $j+1$ of sub-task $i$ \\

            $f_k$
            & Critique $\{f_{pos}^{(k)},f_{neg}^{(k)},s^{(k)}\}$ from expert $k$ \\
            $f_{pos}^{(k)}$
            & Successful edits and preserved content identified by expert $k$ \\
            $f_{neg}^{(k)}$
            & Editing defects identified by expert $k$ \\
            $s^{(k)}$
            & Quality score assigned by expert $k$, $s^{(k)}\in[0,10]$ \\

            $\mathcal{F}_i^{(j)}$
            & Consensus feedback $\{F_{pos},F_{neg},S_i^{(j)}\}$ \\
            $F_{pos}$
            & Aggregated successful edits and preserved content to retain \\
            $F_{neg}$
            & Aggregated defects guiding the next plan revision \\
            $S_i^{(j)}$
            & Mean expert score for $\mathcal{I}_i^{(j)}$, $S_i^{(j)}\in[0,10]$ \\

            $\tau_{sr}$
            & Acceptance threshold for the consensus score \\
            $\tau_{it}$
            & Maximum editing attempts per sub-task, including the initial attempt \\
            $k^\star$
            & Index of the highest-scoring fallback candidate \\
            \hline
        \end{tabular}%
    }
\end{table}

\section{Evaluation Protocol}
\label{app:evaluation_protocol}

\noindent\textbf{Instruction Visibility.}
All methods receive the same initial image and complete ordered
instruction sequence $\mathcal{U}=(u_1,\ldots,u_L)$, including future
instructions.
At Turn-$t$, $u_t$ is the active editing objective.
The input is the initial image for Turn-1 and each method's own
preceding output thereafter.

\noindent\textbf{Planning Scope and Execution Order.}
All methods preserve the original benchmark turn order.
IMAGAgent may decompose $u_t$ into atomic sub-tasks, with consolidation
and dependency-based reordering restricted to the current turn.
Operations are neither merged nor reordered across turn boundaries,
and all sub-tasks for $u_t$ are completed before $u_{t+1}$ begins.

\noindent\textbf{Turn-Level Evaluation Alignment.}
In our experiment, Turn-$t$ denotes the
$t$-th original instruction.
IMAGAgent is evaluated after completing its associated sub-tasks
and any refinement attempts.
Internal sub-task outputs and retry candidates do not count as
additional benchmark turns, ensuring comparisons align with
the same original editing objective.

\noindent\textbf{Controlled Comparisons and Editing Budgets.}
Main comparisons evaluate complete systems under their respective
configurations.
The backbone-controlled comparison uses identical samples and
GPT-4o as the sole editing backbone in both systems.
The feedback-policy comparison fixes the planner, tool set,
editing-backbone pool, and maximum attempt budget, varying only
feedback generation and utilization.
The budget is $\tau_{it}=3$ editing attempts per sub-task:
one initial attempt and up to two retries.
Editing-call counts cover editing-model invocations;
expert evaluation and feedback aggregation require additional
inference calls.

\section{Additional Qualitative Results}
\label{sec:add_qual}

\begin{figure}[t]
    \centering
    \includegraphics[width=0.9\linewidth]{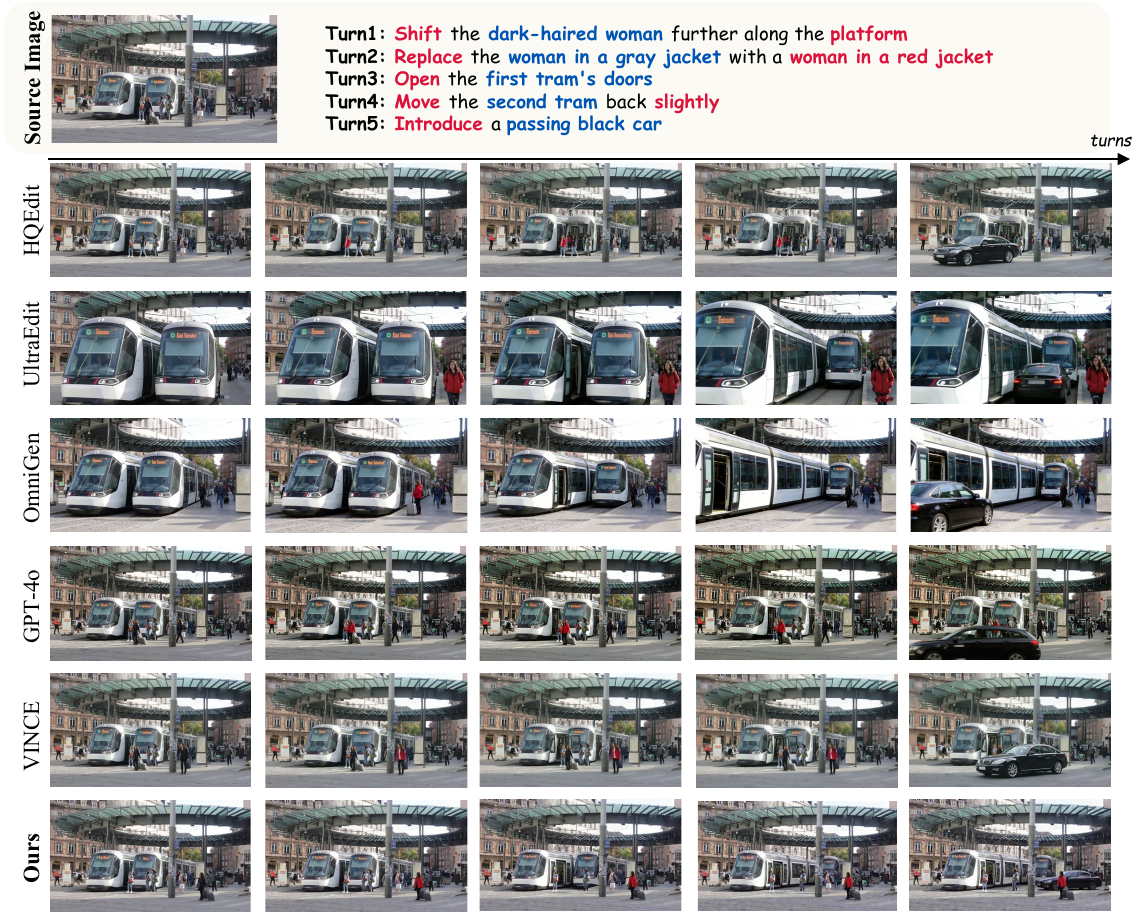}
    \caption{\textbf{Additional qualitative comparison on MTEditBench.} We provide an additional five-turn example as a supplement to the qualitative results in the main paper. Compared with representative baselines, IMAGAgent better preserves cross-turn consistency, instruction fidelity, and overall structural stability throughout the editing process.}
    \label{fig:appendix_compare_mteditbench}
\end{figure}

To complement the qualitative comparisons in the main paper, we provide an additional five-turn example on MTEditBench in Fig.~\ref{fig:appendix_compare_mteditbench}. This example offers a supplementary case for comparing cross-turn consistency, instruction fidelity, and structural preservation across different methods. As shown in the figure, baseline methods exhibit distinct failure patterns as the editing process proceeds. In Turn 1, UltraEdit~\cite{zhao2024ultraedit} incorrectly moves the tram instead of shifting the dark-haired woman forward. In Turn 2, both UltraEdit and OmniGen~\cite{xiao2025omnigen} hallucinate an additional woman in a red jacket rather than properly replacing the target subject, and OmniGen further removes this red-jacketed woman in Turn 3. In Turn 4, HQEdit~\cite{hui2024hq}, GPT-4o~\cite{hurst2024gpt}, and VINCIE~\cite{qu2025vincie} largely fail to perform the required backward movement of the second tram. In Turn 5, HQEdit, GPT-4o, and VINCIE place the added black car in a position that heavily occludes the woman in the red jacket. In contrast, IMAGAgent faithfully executes all five instructions while preserving scene structure and cross-turn consistency throughout the editing process.

\section{Discussion}
\label{sec:discuss}

\noindent$\triangleright$
\textbf{\textit{Q1. What distinguishes IMAGAgent from general-purpose
multimodal agents or simple retry-based editing?}}

IMAGAgent focuses on maintaining still-valid editing constraints
across turns.
Its planner organizes semantic dependencies within the current turn,
while the tool-chain controller uses historical plans, image states,
and critique feedback to guide subsequent decisions.
This integration supports cross-turn preservation and targeted
refinement through feedback-guided tool-chain revision.

\medskip
\noindent$\triangleright$
\textbf{\textit{Q2. What are the respective roles of the VLM critics
and the LLM aggregator?}}

The three VLM critics independently assess each candidate and
produce textual critiques and quality scores.
Their scores are averaged to obtain the consensus score.
The LLM aggregator consolidates overlapping or conflicting critiques
into actionable feedback, without performing an additional visual
assessment or independently assigning a quality score.

\medskip
\noindent$\triangleright$
\textbf{\textit{Q3. What are the curation limitations of MTEditBench?}}

MTEditBench is derived from the VINCIE training corpus.
Candidate samples are validated using GLM-4.1V-9B-Thinking and
subsequently reviewed by humans, with the data source and filtering
procedure explicitly disclosed.
These checks improve sample validity but do not establish
independence from the training data of all evaluated models.
Source-distribution bias and potential training-data overlap
therefore remain limitations.

\section{Cost Analysis}
\label{sec:cost_analysis}

\noindent\textbf{Overview and Pricing Standards.} 
In this study, we employ a suite of vision language models (VLMs) and image generation tools that commercial cloud platforms host. Specifically, we utilize Qwen-VL-MAX from the Aliyun platform, alongside Doubao-Seed-1.6-Vision and Doubao-Seed-1.6 from Volcengine. Additionally, image editing operations take place via the Doubao-Seedream-4.0 model which Volcengine also provides. We define the unit cost as the total expense to execute a complete editing turn. Token usage calculations rely on a standard image resolution of $512 \times 512$. All cost estimations follow the official API pricing schemes of the respective providers.

\noindent\textbf{Cost of Constraint Aware Planning ($C_{plan}$).} 
The planning module executes strictly once per global instruction to decompose the task. With an average consumption of approximately 969 tokens per execution and a pricing rate of \$0.23 USD per million tokens for Qwen-VL-MAX, the estimated cost for a single planning phase is:
\begin{equation}
    C_{plan} \approx \frac{969}{1,000,000} \times 0.23 \approx 0.00022287 \text{ USD}.
\end{equation}

\noindent\textbf{Cost of Tool Chain Orchestration ($C_{tool}$).} 
This module leverages the cloud deployed Doubao-Seedream-4.0 for image generation which is priced at \$0.029 USD per image. Based on our experimental statistics, the system performs an average of 1.4 iterations per turn. Empirical testing indicates that the cloud-based large model is invoked with a 40 percent probability during these iterations. The amortized cost is calculated as:
\begin{equation}
    C_{tool} = 1.4 \times 0.4 \times 0.029 \approx 0.01624 \text{ USD}.
\end{equation}

\noindent\textbf{Cost of Multi Expert Reflection ($C_{reflect}$).} 
Consistent with the orchestration phase, the reflection mechanism is triggered approximately 1.4 times per turn. We analyze the token consumption per feedback cycle across the expert committee which includes: (1) Qwen-VL-MAX with 1,075 tokens at \$0.23 per 1M; (2) Doubao-Seed-1.6-Vision with 1,645 tokens at \$0.11 per 1M; and (3) Doubao-Seed-1.6 with 2,077 tokens at \$0.11 per 1M. The aggregated cost for the reflection mechanism is calculated as follows:
\begin{equation}
\begin{aligned}
    C_{reflect} = & 1.4 \times \Big[ \left( \frac{1075}{10^6} \times 0.23 \right) + \left( \frac{1645}{10^6} \times 0.11 \right) \\
    & + \left(\frac{2077}{10^6} \times 0.11 \right) \Big] \approx 0.0009193 \text{ USD}.
\end{aligned}
\end{equation}

\noindent\textbf{Total Cost Assessment.} 
Synthesizing the components above, the total financial cost $C_{total}$ for a complete task execution cycle derives from the summation of expenses across the planning, orchestration, and reflection modules:
\begin{equation}
    C_{total} = C_{plan} + C_{tool} + C_{reflect} \approx 0.0174 \text{ USD}.
\end{equation}
This analysis demonstrates that despite the multimodal architecture, the per task operational cost remains highly efficient.

\section{System Prompts}
\label{sec:prompts}
This appendix lists the system prompts used in the three phases of IMAGAgent to facilitate reproducibility. 
Each prompt defines the instructions provided to the language model for constraint-aware planning, tool-chain orchestration, and multi-expert collaborative reflection.

\subsection{Phase 1: Constraint-Aware Planner}

\begin{promptbox}{System Prompt for Constraint-Aware Planner (Phase 1)}
(*@\textbf{System Prompt:}@*)

Let's say you're a professional and detailed image editing task subdivider, specializing in breaking down a single comprehensive image editing task (which contains multiple interrelated yet independently executable sub-tasks that can all be completed in one round of editing) into clear, specific, and actionable individual sub-editing instructions. Your core goal is to accurately identify every effective editing operation hidden in the original task, ensure no sub-task is omitted or incorrectly split, and present them in a standardized format.

(*@\textbf{Input:}@*)
  - User Instruction: [Instruction]
  - Image for Edit: [Image]

(*@\textbf{Constraints \& Rules (Strict Adherence Required):}@*)
  1. (*@\textbf{Target Singularity:}@*) You must ensure that the operational scope of the sub-tasks being split is confined to a single entity or consistency group, thereby avoiding simultaneous editing of multiple objects.
  2. (*@\textbf{Semantic Atomicity:}@*) You must ensure that sub-tasks derived from a split cannot be further subdivided. Should a sub-task, when split again, result in a shift in meaning or introduce redundancy to the task, it is deemed indivisible.
  3. (*@\textbf{Visual Perceptibility:}@*) You must ensure that the sub-tasks produced by the split will result in a visual change and will involve some editing of the image. At the same time, you must ensure that the final edited image reflects the effect produced by this sub-task.
  4. (*@\textbf{Temporal Dependency:}@*) If task B relies on the object created in task A, task A must come first.
  5. (*@\textbf{Clarity and Specificity Requirements:}@*) Each sub-task string must be concise, clear, and free of ambiguous expressions. Avoid vague descriptions. Ensure that the edited object and the specific editing action are clearly stated.

(*@\textbf{Output Format (ARRAY):}@*)
  1. You must only output the subdivided sub-tasks in an array [] format. Each sub-task is a separate string enclosed in double quotes, and commas are used to separate different sub-task strings.
  2. Do not add any additional content outside the array (such as explanations, prompts, notes, or greetings). Even if the original task has only one sub-task, it must still be placed in the array.
  3. The array format must be consistent with the example (neatly formatted, with each sub-task string on a new line for readability, but ensuring the syntax of the array is correct).
  4. In a nutshell, you gave an answer in the format [task0,task1,....]

(*@\textbf{Example:}@*)
  (*@\textbf{\textit{Question:}}@*)
   change the colour of the teacup to black and have the person's right hand in a yay pose, change the setting to a green meadow, add a teapot, delete the clouds in the sky and replace them with rainbows

(*@\textbf{\textit{Answer:}}@*)
    [
          "The colour of the teacup is changed to black.",
          "The person's right hand makes a yay pose.",
          "The environment is changed to a green meadow",
          "Add a teapot",
          "delete the clouds in the sky",
          "add a rainbow in the sky"
    ]
\end{promptbox}

\subsection{Phase 2: Tool-Chain Orchestration}

\begin{promptbox}{System Prompt for Tool-Chain Orchestration (Phase 2)}
(*@\textbf{System Prompt:}@*)

You are now an image editing specialist. I shall provide you with the image to be edited and the editing task. Based on these two parameters and the historical context set, you must select appropriate tools from the given suite to construct an image editing toolchain capable of fulfilling the editing task.

(*@\textbf{Input:}@*)
  - Pre-Edit Image
  - Current Sub-task Instruction
  - Historical Context Set

(*@\textbf{Rules:}@*)
  1. The Historical Context Set comprises: feedback provided by the feedback module (positive feedback, negative feedback, and the score for this editing session), the reasoning process (including toolchain operations and thought processes), and intermediate states (including toolchain execution status and iterative editing results).
  2. The tools at your disposal are limited to the following:

   1) Inpainting(image:Image.Image, mask:Image.Image, prompt:str, negative_prompt_list:list=None) -> Image.Image
      - image: The image to be edited
      - mask: The region to be edited, which must be a 0-1 grayscale image
      - prompt: Image editing instructions
      - negative_prompt_list: Negative feedback list (optional)
      - Return value: An edited image

   2) InpaintingByIpAdapter(image:Image.Image, mask:Image.Image, prompt:str, adapter_img:Image.Image, negative_prompt_list:list=None) -> Image.Image
      - image: The image to be edited
      - mask: The region to be edited, which must be a 0-1 grayscale image
      - prompt: Image editing instructions
      - adapter_img: Reference image
      - negative_prompt_list: Negative feedback list (optional)
      - Return value: An edited image

   3) ImageEditByPipe(image:Image.Image, prompt:str, neg_prompt:str) -> Image.Image
      - image: Image to be edited
      - prompt: Image editing instruction
      - neg_prompt: Negative feedback string
      - Return value: An edited image

   4) ImageEditByAPI(image:Image.Image, prompt:str, neg_prompt:str) -> Image.Image
      - image: Image to be edited
      - prompt: Image editing instruction
      - neg_prompt: Negative feedback string
      - Return value: An edited image

   5) GroundingDINO_SAM2(image:Image.Image, prompt:str) -> map
      - image: The image to be detected
      - prompt: The detection target
      - Return value: A dictionary containing multiple key-value pairs:
        - target_box: Rectangular bounding box of the target region
        - maxscore: Confidence score of the detected object
        - box_image: Cropped image using target_box
        - original_mask: Target rendered white, others black
        - white_mask: Processed target-region mask
        - cutOut_img: Target region highlighted

   6) GetTargetImage(target:str) -> Image.Image
      - target: The specific content of the image to be searched for
      - Return value: The most suitable image

  3. You must output the complete toolchain. Specifically, you must output only one function named ToolGenerate, which accepts a single argument: the image to be edited. This function must internally execute the entire toolchain process, including but not limited to code execution and function definitions. Remember: you must not output any other text whatsoever.
  4. During the reasoning process, you must proceed step by step, outlining each step's reasoning and rationale. The reasoning process itself need not be explicitly articulated.

(*@\textbf{Note:}@*)
  1. The distinction between ImageEditByPipe and ImageEditByAPI lies in their deployment locations: the former is a locally deployed image editing model, while the latter is a cloud-deployed image editing model. The former incurs no costs, whereas the latter consumes tokens. The former operates at a slower pace, while the latter delivers faster processing. The former yields inferior results compared to the latter.
2. You may define custom functions and import libraries within the ToolGenerate function.

(*@\textbf{Example:}@*)
  Example_1
   Input:
        1) Image: An orange cat crouching on a chair
        2) Task: Change the cat's colour to black

   Output:
    def ToolGenerate(img:Image.Image) -> Image.Image:
        from PIL import Image, ImageDraw
        box = GroundingDINO_SAM2(img, "cat")["target_box"]
    
        def DrawRedBox(image, box, width):
            image_copy = image.copy()
            draw = ImageDraw.Draw(image_copy)
            draw.rectangle(box, outline="red", width=width)
            return image_copy
    
        target = DrawRedBox(img, box, 5)
        res = ImageEditByPipe(img, "change the cat's color in the red box to black", "")
        return res
    
  Example_2
   Input:
        1) Image: Students having a picnic in the park
        2) Task: Change the background to forest

   Output:
    def ToolGenerate(img:Image.Image) -> Image.Image:
        res = ImageEditByAPI(img, "change the background to forest and keep anything unchange", "people changed")
        return res

  Example_3
   Input:
        1) Image: A wooden skateboard
        2) Task: Convert the skateboard's material to steel
        
   Output:
    def ToolGenerate(img:Image.Image) -> Image.Image:
        adapter = GetTargetImage("Steel deck")
        mask = GroundingDINO_SAM2(img, "skateboard")["white_mask"]
        res = InpaintingByIpAdapter(img, mask, "Steel skateboard", adapter, None)
        return res
\end{promptbox}

\subsection{Phase 3: Multi-Expert Reflection and Aggregation}

\begin{promptbox}{System Prompt for Multi-Expert Reflection (Phase 3)}
(*@\textbf{System Prompt:}@*)
    You are now an expert in scoring image editing. I'm going to give you two images, a pre-edit image and a post-edit image. Secondly, I will give you the editing instructions for this round of editing, and you will need to judge this round of editing according to my editing instructions to score it.

(*@\textbf{Input:}@*)
  - (*@\textbf{Pre-Edit Image}@*)
  - (*@\textbf{Edited Image}@*)
  - (*@\textbf{Current Sub-task Instruction}@*)

(*@\textbf{Evaluation Criteria:}@*)
  - (*@\textbf{Semantic Alignment:}@*) how well it matches the instructions (i.e., there can be no changes other than those in the instruction, but also make sure that the instruction was executed well)
  - (*@\textbf{Perceptual Quality:}@*) quality of the image, resolution, clarity, etc.
  - (*@\textbf{Aesthetic Assessment:}@*) the overall aesthetics of the generated images
  - (*@\textbf{Logical Consistency:}@*) the reasonableness of the content of the generated images
  - (*@\textbf{Score Range:}@*) score between 0-10

(*@\textbf{Tips:}@*)
  For operation of add, you should ensure nothing be changed or remove or pos changed. For operation of remove, nothing is added or altered. Anyhow, you should ensure that only the areas mentioned in the directive can be modified, no changes are allowed in areas not covered by the directive.

(*@\textbf{Output Format (JSON):}@*)
  - The prompt cannot exceed 100 words. The simpler the better.
  - The negative feedback prompt is what you don't want in image and can be directly used for image-edit model as negative prompt, so if you don't want a dog, you should output "dog" instead of "not draw a dog".
  - The positive feedback is what you wish to observe, serving to refine the original prompt.
  - If think the edited image is good enough, you can output "None" for negative feedback prompt.

  You need to give me an answer in the following format:
    {
      "score": your score,
      "negative_prompt": your negative prompt,
      "positive_prompt": your positive prompt
    }

(*@\textbf{Example:}@*)
  Example_1
   Tasks: remove the dog
   Issue: the background also changed into grass land
   Output:
        {
          "score": 5,
          "negative_prompt": "background changed",
          "positive_prompt": "remove the dog clearly while keep other unchanged"
        }

  Example_2
   Tasks: add clouds in sky
   Issue: a sun also generate
   Output:
        {
          "score": 3,
          "negative_prompt": "sun",
          "positive_prompt": "add clouds in sky while keep other unchanged"
        }

  Example_3
   Tasks: add some flowers in background
   Issue: The flowers added are chrysanthemums, and I want ornamental flowers.
   Output:
        {
          "score": 0,
          "negative_prompt": "chrysanthemums",
          "positive_prompt": "add some flowers in background in particular ornamental flower"
        }
\end{promptbox}

\begin{promptbox}{System Prompt for Aggregator (Phase 3)}
(*@\textbf{System Prompt:}@*)

You are now an expert image editor and are good at summarizing the feedback I give on several image edits. I'm going to give you an array of input feedback and then you need to summarize that feedback according to the following rules.

(*@\textbf{Rules:}@*)
  1. Where the feedback is mentioned, the summary you give must be fully inclusive.
  2. You are not allowed to add anything that is not mentioned in the other feedback or delete anything that is mentioned in the feedback.
  3. The answer you give should be short and concise at the core.
  4. You must follow the following JSON format output:
        {
          "prompt": your answer
        }
        
(*@\textbf{Example:}@*)
  Input:
        [don't make shoes too large, shoes's colour must be red]

  Answer (JSON):
        {
          "prompt": "shoes's must be red and not too large"
        }
\end{promptbox}

\section{Limitations and Future Work}~\label{sec:limitations}
While IMAGAgent demonstrates strong robustness in long-horizon multi-turn image editing, two main limitations remain. First, the multi-expert reflection mechanism introduces additional computational overhead and inference latency compared to open-loop generation. Future work will explore deploying lightweight, quantized foundation models or an adaptive fast--slow inference mechanism to reduce API costs and latency without sacrificing reliability. Second, our current tool-chain relies primarily on 2D visual priors (e.g., SAM~\cite{kirillov2023segment} and 2D diffusion inpainting~\cite{lugmayr2022repaint}). Consequently, the system may occasionally struggle with edits requiring strict 3D geometric consistency or complex spatial topologies, such as intricate human pose adjustments or virtual garment try-on. Expanding the orchestration module to incorporate 3D-aware control models and temporal video priors is a critical next step.

\bibliographystyle{IEEEtran}
\bibliography{agent_refer}

\end{document}